\documentclass[useAMS,usenatbib]{mn2e}

\usepackage{graphicx}	
\usepackage{amsmath}	
\usepackage{amssymb}	
\usepackage{soul}
\usepackage{newtxtext,newtxmath}

\title[Eccentric resonant orbits]
{Families of eccentric resonant orbits in galaxy discs: backbones for bars and spirals}
\author[C. Struck] 
{Curtis Struck \thanks{E-mail: curt@iastate.edu} \\
Department of Physics and Astronomy, Iowa State University, Ames, IA, 50011 USA}
\usepackage{amssymb}
\usepackage{amsmath}
\usepackage{graphicx}
\usepackage{multirow} 
\def\aap{{ A\&A}}

\def\aj{{AJ}}

\def\apj{{ApJ}}
\def\apjl{{ApJL}}

\def\mnras{{MNRAS}}

\begin{document}
\date{\today}

\pagerange{\pageref{firstpage}--\pageref{lastpage}} \pubyear{0000}

\maketitle

\label{firstpage}
\begin{abstract}

It is widely believed that resonant orbits play an important role in formation and evolution of bars and large-scale spirals in galaxy discs. These resonant orbits have been studied in a number of specific potentials, often with an imposed bar component. In this paper I show that families of resonant (e.g., two-dimensional $x_1$) orbits of differing eccentricities can be excited at a common pattern speed, in a variety of axisymmetric potentials. These families only exist over finite ranges of frequency in most of these potentials. Populations of such resonant eccentric orbits (REOs) can provide the backbone of both bars and spirals. At each frequency in the allowed range there is a maximum eccentricity, beyond which the REOs generically become quasi-stable (or `sticky'), then unstable (or chaotic), as the eccentricity increases, at values that depend on the potential and the orbit frequency. Sticky and chaotic orbits have been extensively studied recently with invariant/unstable manifolds in a variety of phase planes, but it is found that studying them as a function of eccentricity and pattern speed provides a particularly useful framework for classifying them and their stability transitions.  The characteristics of these orbit families depend on the galaxy potential and the pattern speed, and as backbones of bars and spirals can help understand a number of observed or predicted regularities. These include: the size and speed of bars in different potentials, the range of pattern speeds and windup rates in spirals within galaxy discs, and constraints wave growth.

\end{abstract}


\section{Introduction and background}

\subsection{Resonant orbits in bars and spirals}
\subsubsection{Bar development}  

\citet{lyndenbell79}, also see \citet{lyndenbell72}, conceived a theory of bar formation or enhancement based on resonant orbits. Lynden-Bell's work showed how the gravity of eccentric resonant orbits could essentially trap near-resonant orbits, and build-up the bar potential, leading to more orbit trapping. Instead of precessing away from the resonant orbit, in this picture captured orbits will precess backwards and forwards around the resonant orbit. This might be seem a slow process of bar formation, because a nascent bar consisting of a small ensemble of orbits at an ILR will not generally have strong self-gravity to capture orbits. On the other hand, in a rising rotation curve region resonant and near resonant orbits can be excited over a range of radii, and could make a much stronger proto-bar potential. As noted by \citet{lyndenbell72}, this idea of using a solid-body rotation curve or harmonic potential to essentially turn the ILR into an extensive region of resonant orbits dates to Lindblad's early work (\citealt{lindblad58} as quoted in \citealt{lyndenbell72}). The consensus theory of bars holds that they have a backbone of nested orbits in the pattern frame in the inner disc region with such a potential (e.g., see the reviews of \citet{athanassoula13, sellwood14}). In the wave picture, this backbone is the standing wave. \citet{polyachenko04}, also see \citet{polyachenkos04}, has proposed a variant of Lindblad theory in which bars and spirals are formed from unstable slow modes, with a range of differential precession rates. 
  
A primary goal of this paper is to extend these ideas, and especially the Lynden-Bell theory, to a wider range of disc potentials. As will be shown below, in a wide range of potentials, there exist extended radial ranges with eccentric orbits that are nearly in resonance with a single pattern speed. This idea is based on the work of \citet{struck15b}, which showed that the locations of ILRs depends on eccentricity as well as pattern frequency. Ensembles can be made up of selected resonant eccentric orbits (REOs) with eccentricities that vary slowly with size. The excitation of these orbits could lead to more robust bar formation in more cases than previously considered.
  
\subsubsection{Spirals}

There are a wide variety of spirals in disc galaxies, including long, open or tightly wound `grand designs', numerous short, `flocculents', and spirals originating at the ends of bars. One theoretical approach to understanding grand design spirals is the global modes theory (\citealt{bertin00, dobbs14}), in which it is proposed that a small number of large scale unstable wave modes can grow to saturation, forming long-lived standing waves. This is in contrast to Lin-Shu type waves which are locally unstable, though perhaps over an extended radial range, and are amenable to linear perturbation analysis. Global modes are more difficult to analyze, and much remains to understand about their growth, saturation and persistence. We will see below that in some potentials a backbone of REOs could be an important part of such modes. 

 \citet{kalnajs73} demonstrated how a kinematic spiral wave can be generated from near-overlapping elliptical orbits. He also emphasized how, with the differential precession in a typical disc potential, such a wave would wind up rapidly. In later sections, we will expand this idea, to show that REO ensembles could form the backbones of a wide variety of spirals. These include some that wind up quite slowly. These backbone orbits can be quite eccentric, extend over a significant radial range, but extend over a small range of precession frequency, persisting longer. Like bars, kinematic spirals with REO backbones will likely capture near resonant orbits by the LBK process. It has previously been suggested that sticky REOs (described below) support and enhance the longevity of spirals \citep{harsoula11, harsoula21}. In other cases, transient, local resonant orbit crowding could produce dynamic flocculent waves.  

\subsection{Finding REOs and sticky eccentric orbits}

From the 1970s \citet{contopoulos75, contopoulos80} began the study and classification of the resonant orbits near ILRs idealized in the LBK theory. \citet{contopoulos89} reviewed this work and \citet{athanassoula92} extended it to a four parameter family of Ferrers type bar potentials. This body of work established the idea of nested $x_1$ orbits in near harmonic potentials near an ILR as the backbone of bars, as well as introducing a variety of other orbit families as likely contributors to the bar potential (e.g., \citealt{patsis97}). The early studies of bar orbits were usually in two dimensions (and with fixed potentials), but were later extended to three dimensions, often in the context of understanding boxy bars \citep{pfenniger91, skokos02a, skokos02b, patsis03, patsis14}.  Numerical studies also advanced to broader classes of orbits, including chaotic orbits, now believed to be an important part of the trapped orbit population (e.g., \citealt{patsis06, voglis07, harsoula09, manos11, machado16}). Recent work \citep{gajda16, patsis19} now suggests that nested $x_1$ orbits are not the dominant type, but rather a distinct minority. LBK theory does not require them to dominate. 

Another important class of orbits has emerged in recent decades - `sticky' orbits, which we define as nearly stable for some number of orbits, but which ultimately drift away from a resonant closed form. Sticky orbits have been studied extensively in the context of regions between stable and unstable manifolds in various phase spaces, e.g.,  \citet{habib97, terzic04, katsanikas13, zotos17}. (Note that the \citet{habib97, terzic04} papers, while evidently introducing the term into astrophysics, use a different, but related definition of stickiness. \citet{sellwoodwilk93} describe confined, stochastic orbits without using the term.) I demonstrate below that the transitions through stability-stickiness-instability generally occur continuously, at a fixed wave pattern speed, with increasing orbital eccentricity in many potentials. The complexity of stable/unstable manifolds in phase spaces explored in the references above shows that the $x_1$ transitions to stickiness described below are not unique in multi-component potentials. However, such sequences do appear to be unique and complete for transitions from $x_1$ REOs to sticky orbits in simple monotonic potentials. This simple and well-defined transition sequence may serve as a useful prototype.

In this paper I will argue that the view that nested resonant orbits resulting from harmonic type potentials are the only backbones in bars is too restrictive. One aspect of this argument is that eccentric orbits have an eccentricity dependence in their generalized Kepler Third Law, and so have eccentricity dependent  $(\Omega - \kappa/2)$ curves in most galaxy potentials \citep{struck15b}. This means that,  in a variety of potentials, at a given pattern speed, there may exist a finite range of eccentricities, and (at least partially) nested REOs of different sizes (i.e., semi-major axes). This eccentricity range of REOs pinches off to zero at high orbital frequencies, though the exact value of the termination frequency is potential dependent. Several examples are described below, and the boundaries of the REO zone will be explored in various potentials. This behavior is not captured by the epicyclic approximation.

The existence of REOs over finite eccentricity ranges in non-harmonic potentials helps us to better understand why bars often extend beyond the quasi-harmonic part of a disc potential, e.g., into flat rotation curve regions. Trivially, very eccentric REOs may simply extend well beyond the harmonic region where they form, but the fact that nested REOs can exist in non-harmonic potentials provides a more robust explanation of the phenomenon. 

The range of REOs found in various potentials also helps understand the effects of central concentration on bar development, which has been studied in a number of publications \citep{pfenniger90, norman96, das03, shen04}. There has been some debate in the literature on when or how exactly the growth of central concentrations weakens bars, and how massive the central concentration must be (e.g., \citealt{athanassoula05, bournaud05, debattista06, wheeler23}). We will see in the following sections that while REOs can persist in concentrated potentials, there are more limitations on the possible nested orbit backbones.

This paper has two general parts; the first (Sec. 2 and 3) focuses on REO orbits, and the second (Sec. 4-6) on kinematic bar/spiral backbones made up of these orbits. Specifically, Section 2 briefly reviews the formalism of a precessing ellipse approximation for orbits in galaxy potentials focusing on an example potential consisting of an exponential disc plus a central point mass. The work described in this section is analytic. Section 3 includes a broader examination of numerically computed REOs in both the exponential plus point mass potential and a range of power-law potentials. In Section 4 the analytic orbit approximations are used to examine a wide range of orbit ensemble evolutions, which can provide the backbones of kinematic bars and spirals. In Section 5 the broader ramifications of the results to theory and observations of bars, large-scale spirals are discussed. Section 6 provides a brief summary and overview.

\section{Theoretical formalism and first examples}

In this section the approximation formalism to be used below will be described, and a first example potential considered. Textbook disc potentials typically have multiple components (disc, halo and perhaps a bulge), with at least one Inner Lindblad Resonance ILR), a corotation resonance (CR) and an Outer Linblad Resonance (OLR), over a range of possible pattern speeds (see e.g., \citealt{sparke00}, Ch. 5; \citealt{bt08}). A solid-body rotation curve is an example of where there can be continuous ILRs over some radial range with a pattern frequency equaling the rotation frequency. Other mass distributions in the inner parts of galaxy discs can give rotation curves similar to solid body ones over finite radial ranges, and thus, have a nearly constant values of $\Omega - \kappa/2$ over that range. This can give rise to nested REOs over a limited radial range to support a bar. Beginning in this section we will consider several such potentials.

\subsection{Exponential plus point-mass equation of motion} 

We begin with a fairly simple example potential, characterized by a central concentration plus an exponential disc mass distribution. The central concentration is represented by a point-mass, and could be either a supermassive black hole or a compact bulge. The exponential mass distribution can form quite quickly in a galaxy disc, e.g., \citet{wilman20, robertson23}. In the present case we assume that this exponential plus point-mass potential dominates any additional dark matter contribution throughout the disc. The ratio of the masses of the point mass and exponential components, e.g., within one exponential scale length, is viewed as an adjustable parameter.

In this paper we will limit consideration to planar orbits of massless particles (stars) in fixed, symmetric potentials. The radial equation of motion for a particle is then,  

\begin{equation}
\label{eq1}
\frac{d^2r}{dt^2} =  -g_r + \frac{v^2_\phi}{r},\\
\end{equation}

\noindent
where $r$ is the particle radius, $v_\phi$ its azimuthal velocity in the disc plane, and $g_r$ is the magnitude of the gravitational acceleration. The azimuthal motion is determined by the conservation of angular momentum. 

In the case of an exponential disc with a central point mass, the magnitude of the  acceleration $-g_r$ can be written,

\begin{multline}
\label{eq2}
g_r = \frac{GM_\epsilon}{H\epsilon}\ 
\frac{\epsilon}{r} 
\left[ 1 - \left( 1 + \frac{r}{\epsilon} \right) e^{-\frac{r}{\epsilon}} \right]
+ \frac{GM_B}{\epsilon^2}\ \left( \frac{\epsilon}{r} \right)^2 \\
= c_D u \left[ 1 - \left( 1 + \frac{1}{u} \right) e^{-\frac{1}{u}} \right]
+ \chi c_D u^2, \\
\text{with},\ \  c_D = \frac{GM_\epsilon}{H\epsilon},\ \ \chi = \frac{M_B H}{M_\epsilon\epsilon}, \ \ u = \frac{\epsilon}{r},
\end{multline}.

\noindent where the exponential scale length $\epsilon$ is adopted as the scale radius, and the mass $M_\epsilon$ of the exponential disc component contained within a scale length, is adopted as the scale mass. $M_B$ is the mass of the point-mass, and $H$ is an assumed vertical disc thickness. Examples of the circular rotation frequencies and relatively flat $(\Omega - \kappa/2)$ curves with this acceleration, and with various parameter values, are shown in Fig. 1 and discussed below. However, before studying orbits in these potentials we consider an eccentricity effect.

 \subsection{Eccentricity dependent resonances}
 
 In addition to the case of rotation curves that give nearly flat $(\Omega - \kappa/2)$curves over some range of radii, there is a second way to get extended resonance regions. This is based on the fact, noted above, that at a given radius the resonant frequency has a modest dependence on orbital eccentricity. More eccentric orbits have lower inner Lindblad resonant frequencies than circular orbits, as shown in Fig. 1 and discussed in \citet{struck15b}. Alternately, for a given pattern speed, there may be a range in radius, within which there is a resonant orbit of specific eccentricity centered at each radius. (The approximate analytic version of this region was called the Lindblad Zone in \citealt{struck15b}). Thus, resonant orbits could be excited throughout this finite radial range, though they would span a range of eccentricities. 
 
 This phenomenon can be studied via orbital solutions to equation \eqref{eq1} above. We will consider numerical solutions in the following section. Approximate orbit solutions can also be obtained from the precessing, power ellipse, or p-ellipse, approximation, developed in \citet{struck06}. In this approximation we assume orbital solutions of the form,

\begin{equation}
\label{eq3}
u = \frac{1}{p} \left[ 1 +
e \cos \left( m{\phi} \right) 
\right]^{\frac{1}{2} + \delta}.
\end{equation}

\noindent
which describe precessing, ellipse-like orbits. These orbits depend on three parameters: the eccentricity, $e$,  the semi-latus rectum, $p$, and the factor $m$, which is the ratio of precession and orbital frequencies. Henceforth we will use dimensionless variables scaled to the length $\epsilon$ and the mass $M_\epsilon$ as described in \citet{struck06}. This approximation was originally developed for power-law potentials, where it is quite accurate over the applicable range of radial powers $\delta$. E.g., values of $\delta = \frac{1}{2}, 0$, and $-1$ correspond to the Kepler, flat rotation curve, and solid-body potentials, respectively. The approximation is also reasonably accurate for other continuous potentials, like the one adopted here.

The eccentricity dependence of the precession, given by $m(e)$, is of particular interest here. Simple analytic expressions were given in \citet{struck15a} to approximate the effect in power-law potentials. As a rough average, to obtain approximate estimates of the effect, we adopt a correction factor to $m_1$ equal to that of the $\delta = -0.3$ power-law potential, which has a modestly rising rotation curve. This correction factor yields the eccentricity dependent resonances, and Fig. 1 provides several illustrations. 

\begin{figure}
\centerline{
\includegraphics[scale=0.72]{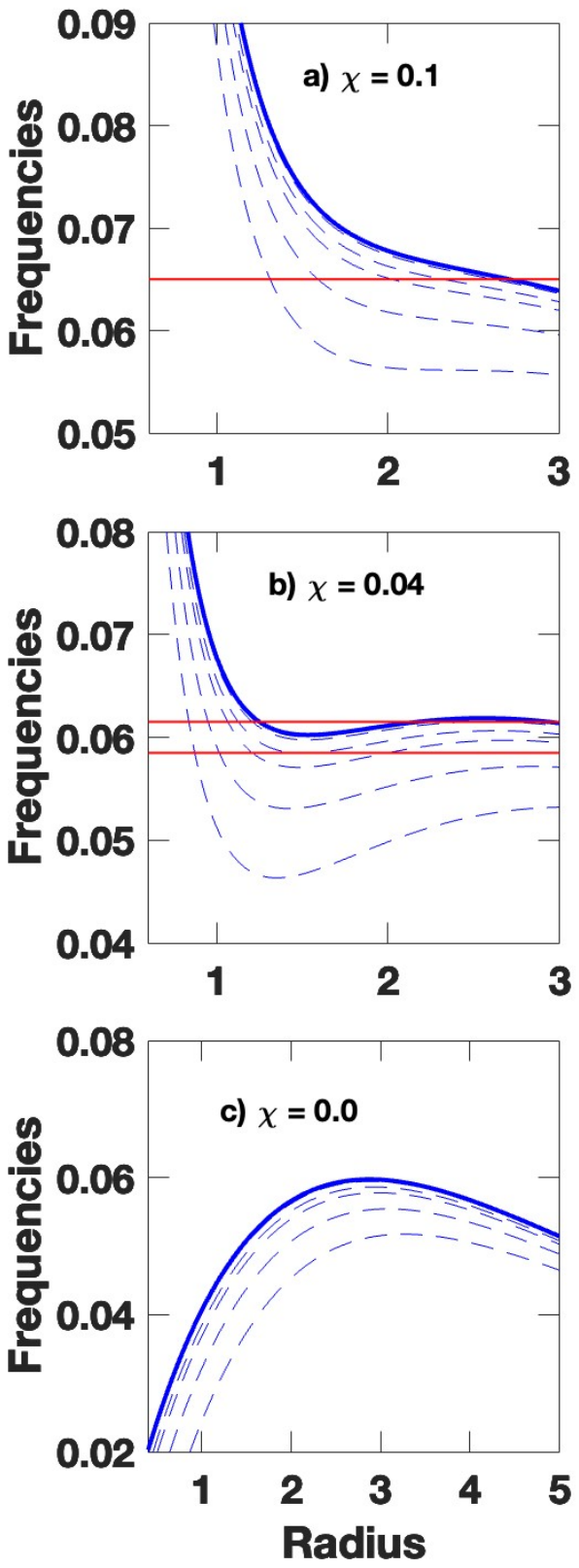}}
\caption{Dimensionless orbital frequency versus disc radius (in units of $\epsilon$) in three cases in an exponential plus point mass potential, described in the text. In each case the blue, solid curve shows the $\Omega - \kappa/2$ locus. Red horizontal lines show representative wave pattern frequencies. $\chi$ is the relative strength of the point-mass and exponential potential components, or a concentration factor, defined in equation (2). Zoomed views of the near constant ILR frequency regions are shown in each case. The dashed lines show approximate $\Omega - \kappa(\epsilon)/2$ curves with generalized epicyclic frequencies, from equations (5) and (6)  at eccentricities of $\epsilon = 0.3, 0.5, 0.8, 0.9, 0.98, 0.999$ from top to bottom. The curves with the smallest eccentricities overlap at the figure resolution. See text for discussion.} 
\label{fig:fig1}
\end{figure}

Each panel in Fig. 1 corresponds to a particular example of the potential of equation \eqref{eq2} with a specific value of the bulge-to-disc parameter $\chi$, varying from a small bulge case in the top row to a larger one in the bottom. The $(\Omega - \kappa/2)$  curves are shown as a blue solid curves. Red horizontal lines are illustrative pattern speeds. Beneath the $(\Omega - \kappa/2)$ curves in each panel are five dashed curves for the eccentricity dependent $(1 - m(e)/2)\Omega$ curves, where the eccentricity dependent factor $m(e)\Omega$ replaces the usual $\kappa$ factor. These dashed curves are computed at eccentricities of $(0.3, 0.5, 0.8, 0.9, 0.98, 0.999)$. Since the eccentricity corrections are small, these dashed curves adhere closely to the $(e=0)$ solid $(\Omega - \kappa/2)$ curve.  

\subsection{Analytically predicted resonant eccentric orbits}

When a pattern speed line intersects one of the resonant curves in Fig. 1, we satisfy the resonance relation, 

\begin{equation}
\label{eq6}
\left[ 1 \pm \frac{m(e)}{n_r} \right] \Omega
= {\Omega}_p,
\end{equation}

\noindent where $\Omega$ is the circular frequency, $\Omega_p$ is the pattern frequency, and $n_r$ is the mode number (e.g., $n_r = 2$ for the curves in Figure 1). Here $m(e)\Omega$ replaces the epicyclic frequency $\kappa$ in the conventional form of this equation. The two versions of the equation are identical when $e=0$, but equation \eqref{eq6} is a generalization to nonzero values of the eccentricity. 

As a first example of what these figures can tell us, in the upper panel of Fig. 1 ($\chi = 0.1$ case), an intersection of the pattern speed line with the second lowest dashed curve represents a closed, symmetric elliptical orbit of eccentricity $e = 0.98$ in a reference frame rotating at the pattern speed. This orbit is nominally centred at the radius of intersection ($r \approx 1.7$). 

As we move to the right along this pattern speed line, it will intersect curves of progressively lower eccentricities until it reaches the solid ($e=0$) $(\Omega - \kappa/2)$ curve. Thus, in this approximation, closed orbits with a range of eccentricities in this pattern may exist over a significant range of radius (nominally about $r = 1.7-2.7$). We will consider examples of such orbits in the following section. Such orbits and the bars and spiral waves that can be made from them in the case of power-law potentials were described in \citet{struck15b}. Most of the examples in that paper focused on a flat rotation curve (FRC) potential. In the present potential the ILR resonance curves are flatter, and in many cases the eccentric (dashed) curves deviate farther from their parent $e=0$ (solid) curve. This means that closed resonant orbits can be excited over a larger range of radii. 

There are several other differences between the resonant orbit populations implied by Fig. 1 and those in FRC-like potentials. Consider a second example of the top pattern speed line in the middle panel of Fig. 1 ($\chi = 0.04$ case). On the left side of this plot ($r < 1.0$) we evidently have a situation similar to the previous example, albeit more compressed in the radial range. Then there is a region where the pattern speed line exceeds the value on the $(\Omega - \kappa/2)$ curve, so we expect no REOs there. And finally, on the right part in the plot ($r > 2$) resonant orbits are possible again with increasing eccentricity as the radius increases. Possibly two bars, an inner and outer one, with the same pattern speed, could form around eccentric closed orbits in this potential, with a non-resonant intermediate zone between them. 

A third example highlights the differences with monotonic potentials. This time we will consider the lower pattern speed curve in the middle panel of Fig. 1 ($\chi = 0.04$). The interesting feature of this example is that for relatively large radii, e.g., $r > 2$, the resonance curves are quite flat. As a result, this pattern speed line in particular, nearly tracks a single resonance curve with $e \approx 0.95$ over a large range of radii. This suggests that  there can be many nested, closed orbits in the pattern frame that could provide a strong backbone for a bar component. 

With many closed orbits, and especially when these are nested and of similar eccentricity, \citet{kalnajs73} showed how a radially dependent twist of those orbits can produce spiral waves. Closed orbits with similar precession frequencies could support a long-lived (or `quasi-steady') spiral feature, with a long windup timescale. Stars on the closed, backbone orbits would spend a large fraction of their orbital periods within or near the spiral.

More generally, Fig. 1 shows that finite ranges of closed resonant orbits can exist over a significant range of parameter values in the adopted potential. Since this potential is a rather generic realization of the combination of a falling rotation curve central potential, and a slightly rising (then flattening) rotation curve disc potential, we can expect this result to be quite general. It can be viewed as an extension to the result in Lynden-Bell's theory \citep{lyndenbell79} that bars should form in rising rotation curve regions. 

The caveat, however, is that the results above and Fig. 1 are based on some rough approximations. This includes the use of a rough approximation for $m(e)$ in the resonant curves in the figure. Moreover, in regions where the eccentricities of the resonant curves are changing fairly rapidly with radius along a pattern speed line it is not clear that an eccentric orbit, which would traverse a significant range of radii, is truly closed (or stable). In the next section we will investigate these issues with explicit numerical integrations of orbits in cases like the examples above.

\section{Example bar/spiral supporting families of REOs}

In the previous section the eccentricity dependent resonance conditions were used to predict the existence of closed elliptical orbits at various pattern speeds. As noted above the procedure was approximate, but is useful for obtaining a qualitative understanding of the location of the REOs, as in Fig. 1. In this section numerical integration of the equations of motion are used to find examples of the predicted REOs. The primary goal here is to look at the structure of orbits at different pattern frequencies in a wide range of potentials, including those described in the previous section. We find, for example, the range of radii and the maximum eccentricities of stable REOs in these potentials. We also see how the maximum allowed pattern frequency for stable REOs varies between these potentials. Note that for the purposes of this paper stable, resonant orbits are defined as closed, non-evolving orbits in the pattern frame. Nearly closed orbits, or orbits that are not closed, but remain close to an average closed orbit at all times (like the green orbit in the top panel of Fig. 2) will be referred to as quasi-stable or near stable.

The examples in this paper will also be limited to $n_r = 2$ REO families located in the vicinity of the $\Omega \pm \kappa/2$ curves. There are other, similar families, especially those with $n_r > 2$, but these will not be considered here.

In all the examples that follow, the search for REOs was carried out by choosing a maximum radius of the orbit, and successively approximating the angular momentum (or azimuthal velocity) corresponding to the symmetric closed orbit with that maximum radius in the pattern frame. This was done by trial-and-error. 

\subsection{Example 1: Exponential disc with a relatively large central mass}

\begin{figure}
\centerline{
\includegraphics[scale=0.78]{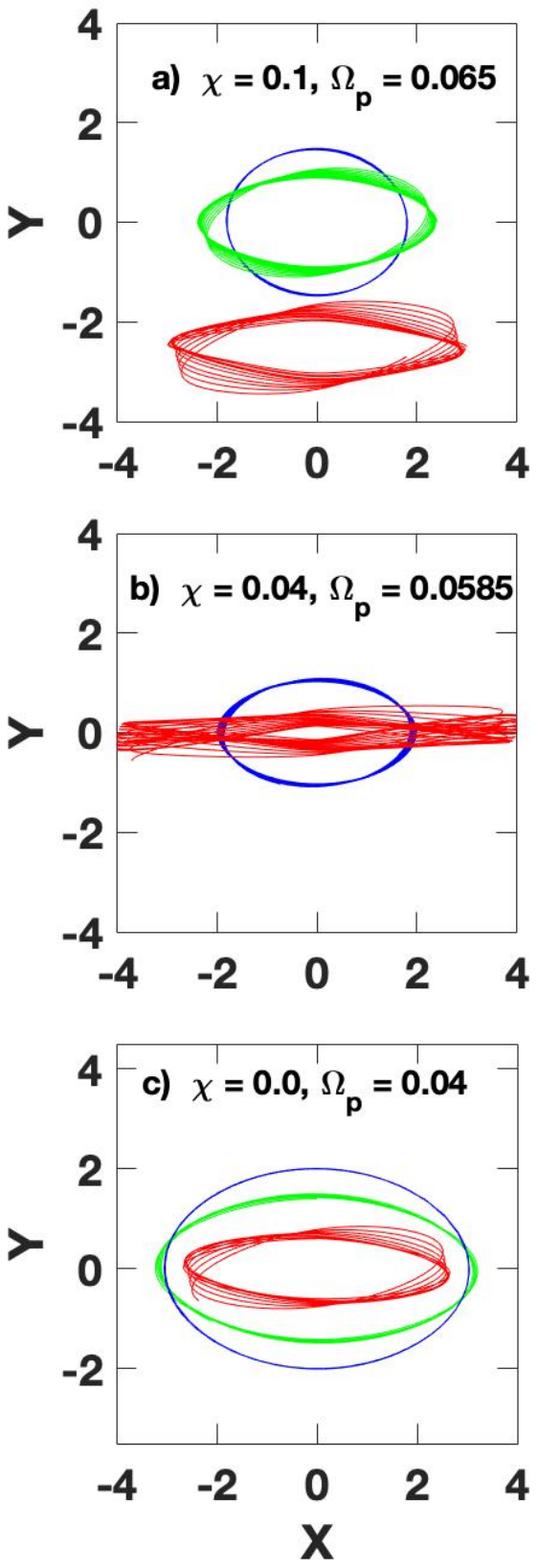}}
\caption{Sample resonant and near resonant orbits in three cases of the exponential plus point-mass potentials. Panels a) - c) show a sequence of decreasing importance of the point-mass component, quantified by the $\chi$ paramenter. Pattern frequencies are chosen to correspond to cases of interest in Fig. 1. Note the red orbit in the top panel is displaced downward by 2.5 units for clarity. See text for details.} 
\label{fig:fig2}
\end{figure}

The top panel of Fig. 2 (Fig. 2a) shows several closed orbits in the potential shown in the top row of Fig. 1 ($\chi = 0.1$, a relatively large central mass). These orbits are shown in the pattern frame corresponding to the red line of the upper panel of Fig. 1 ($\Omega_p = 0.065$). Fig. 2a shows a stable REO (blue, thin curve) and nearly stable REO (green thicker curve) centered on the origin. The stability of these moderate eccentricity orbits suggests that REO orbits exist with typical orbital radii between about $1.0$ and $2.5$ units. The green, thicker orbit is in fact `sticky' since it appears stable for a number of circuits, but is beginning to drift away in the last few circuits shown. A more eccentric and less stable orbit is shown in red, and displaced downward by 2.5 units for clarity. Since all three orbits were computed for the same duration, comparison of the last two show that increasingly eccentric orbits
become increasingly unstable, and drift away from a closed form. Highly eccentric orbits are unstable from the outset and never close. Note that the size difference between the least eccentric unstable orbit and much more unstable orbits is small.

The most eccentric orbits in Fig. 2a also have the smallest minimum radii. Since the more eccentric orbits have large maximum radii they cross the more circular orbits. This suggests that if interstellar gas elements were driven onto such orbits, over a range of radii and eccentricities, there would be cloud collisions and shocks at orbit intersections, see \citet{athanassoula92b}. These dissipative processes could produce a more nearly laminar flow at some weighted average eccentricity in a bar.  

\subsection{Example 2: Intermediate exponential/point-mass case}

The middle row of Fig. 1 shows a case with an exponential disc and an intermediate point mass, which has a minimum in the $(\Omega - \kappa/2)$ curves. As discussed above, at certain pattern speeds (e.g., $\Omega_p \simeq 0.062$) there exist two regions where we might expect to find REOs. These are separated by a region where we expect no REOs at the given pattern speed. This is a complex example, and it is not clear how accurate the linear theory these curves are based on is.

 Fig. 2b shows a numerically calculated REO, the blue curve, with a pattern frequency in the interesting region of the
 $\Omega - \kappa/2$ curve. A second, flatter orbit (red curve) shown is nearly stable, though visibly thick. Orbits only slightly more eccentric than the (blue) REO are quite unstable, but at higher eccentricities they become marginally stable like the flat, red curve. This seems to be the realization of the unstable intermediate region predicted in the middle row of Fig. 1. 

\begin{figure}
\label{fig3}
\centerline{
\includegraphics[scale=0.64]{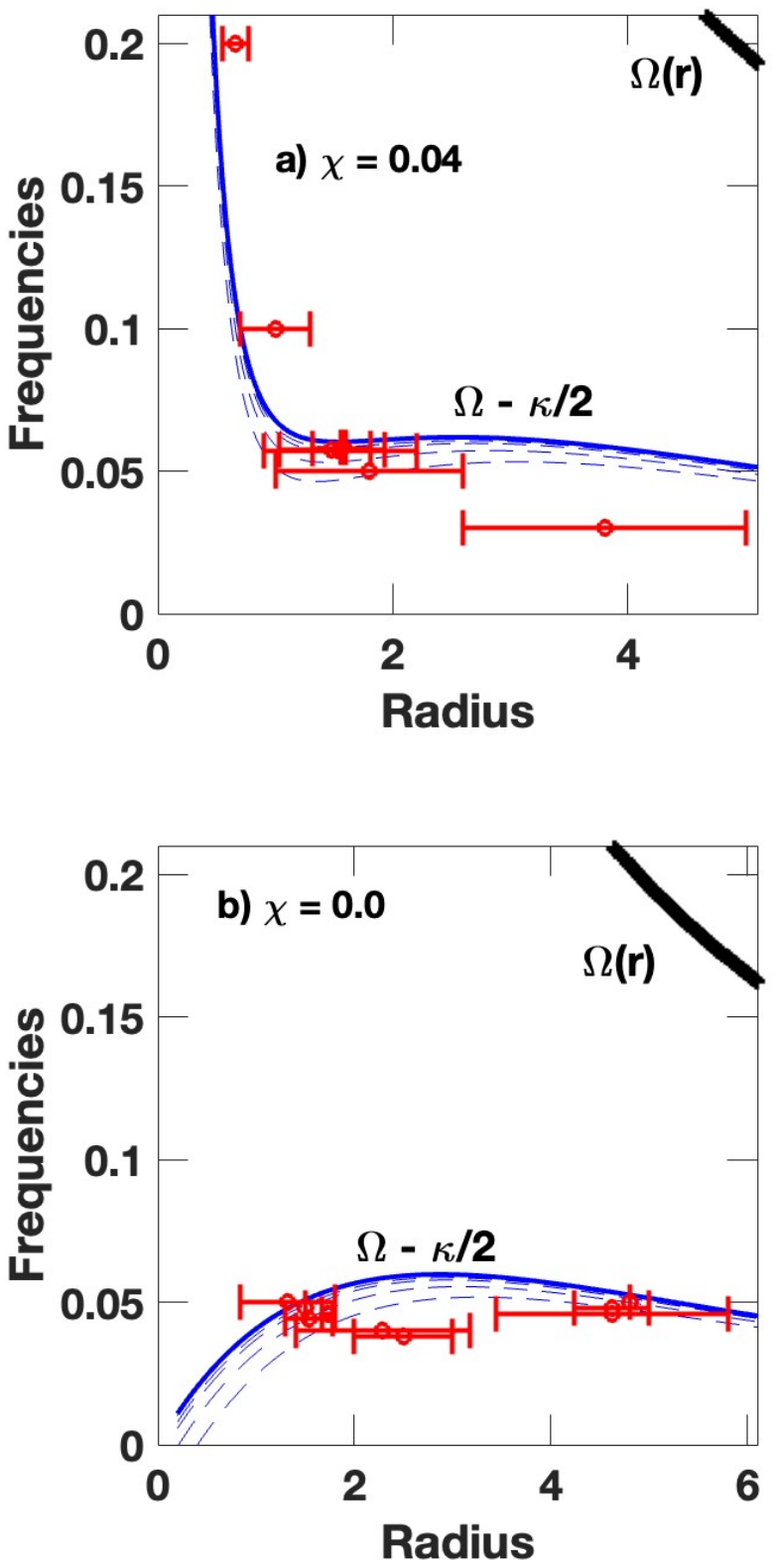}}
\caption{Orbital frequency versus radius for two values of the concentration parameter $\chi$ of the exponential plus point-mass potential as in Fig. 1. The blue, lower solid curve shows the $\Omega - \kappa/2$ locus, the black, upper thick curve the $\Omega(r)$ curve, and the dashed curves show generalized $\Omega - \kappa(\epsilon)/2$ curves with the same eccentricities as Fig. 1. Red, horizontal error bars show the radial extent of a number of repesentative REOs, as derived from numerical orbit calculations. See text for details.} 
\label{fig:fig3}
\end{figure}

Fig. 3a shows another view of the frequencies in the intermediate case ($\chi = 0.04$) of the middle row of Fig. 1. Fig. 3a additionally gives examples of the full radial extent of some REO orbits at a variety of frequencies, each shown as horizontal line segments, with error-bar type ends for the minumum and maximum radii of the orbits, and dots showing the midpoints. The cluster of these error-bar segments at a frequency of about $\Omega_p = 0.0585$ (as in Fig. 2b) contains four orbits with a range of eccentricities, and all centered near a radius of 1.55 units. These four orbits with the same frequency, but very different eccentricities, all have about the same midpoint radii. Although not shown, REOs with a range of eccentricities exist at other frequencies as well. 

The other error-bar segments in Fig. 3a show the extent of maximally eccentric, stable REOs, i.e. non-wobbly orbits, at various frequencies: $\Omega_p = 0.035, 0.050, 0.10$, and $0.20$. The determination of the maximum value of the eccentricity for a stable periodic orbit was done by a visual examination of individual orbits near the critical value at each pattern speed. Thus, it is not a highly accurate determination, but because of the fairly rapid transition from stability to instability it is be reasonably so. 

Note that the most eccentric REOs are found well below the blue solid curve in Fig. 3a. This is what we would expect from the distribution of the dashed blue curves, with eccentricity increasing downwards, and the results for power-law potentials studied in \citet{struck15b}. Note that the value of the eccentricities of the error-bar segments do not match the values of the closest dashed curves very precisely; only the pattern frequency is plotted and this and similar figures, not the total stellar orbital frequency. In this more complex potential, the theory behind the dashed curves, e.g., the adopted form of the function $m(\epsilon)$ in equation \eqref{eq6}, is only approximate. Fig. 3 of \citet{struck15b} also shows the onset of wobbly orbits above a critical eccentricity in a flat rotation curve potential. 

The four clustered error-bars in Fig. 3a illustrate another point. At a given pattern frequency REO orbits exist with all eccentricities less than the maximal periodic one.  

At the highest frequencies shown, the most eccentric REO is not very eccentric. At somewhat higher frequencies the maximum eccentricity goes to zero, and no eccentric orbits close in the pattern frame. 

In the potentials of Figs. 1 and 3 the $\Omega(r)$ (corotation) curve lies well above the regions of the REOs shown. None of the error-bar REOs shown in Fig. 3a extend at all close to corotation at the given pattern speeds. Bars are usually classified as slow or fast based on the ratio ($\mathcal{R}$) of corotation radius at the given pattern speed to the maximum size of the bar. By this criterion, a bar supported by any of the REOs in this family, and of comparable size, would be classed as very slow. In this potential it is not possible to have a fast bar supported by a REO. Wobbly orbits do extend to higher eccentricities, and somewhat greater lengths, but numerical calculations show that the extra extension is not enough to contradict this result. 

Finally, note that low eccentricity REOs do exist near the OLR, which is well above corotation, and also not shown in Figs. 1 and 3. These seem unlikely to support bars, but might be associated with elliptical or oval discs.

\subsection{Example 3: Pure exponential case}

Fig. 2c shows some sample orbits in a exponential potential with no point mass component. In this case too we see the sequence of stable, nearly stable, and sticky, but ultimately unstable orbits, with increasing eccentricity. As we will see below this seems to be a universal sequence. 

Fig. 3b shows the radial extent of a number of REOs like those in Fig. 3a. As might be expected from the curvature of the resonance curves, this is also a complicated case. The orbits shown were chosen arbitrarily, but illustrate that a wide range of REOs could be excited over a narrow pattern frequency range. Sequences of nested, non-crossing orbits for bar support could be found readily in this variety of orbits.

The complexity of these exponential plus point mass potential cases may obscure some trends, and leave questions of which results are unique to these potentials. To provide clarity we conclude this section with a consideration of simple power-law potentials varying concentration.

\subsection{Example 4: Flat rotation curve case, }

REO orbits in this case, where $\delta = 0.0$ in equation (3), were already considered in \citet{struck15b}. But as noted in the previous subsection, the analytic and numerical results were not compared in detail. Part of that comparison is illustrated in Fig. 4, where the $\Omega - \kappa/2$ curve, the $\Omega$ curve, and the $\Omega + \kappa/2$ curve are shown, and also error-bars for the for the maximally eccentric stable REOs at various pattern frequencies. The latter were determined numerically as in the previous subsections. 

\begin{figure}
\centerline{
\includegraphics[scale=0.39]{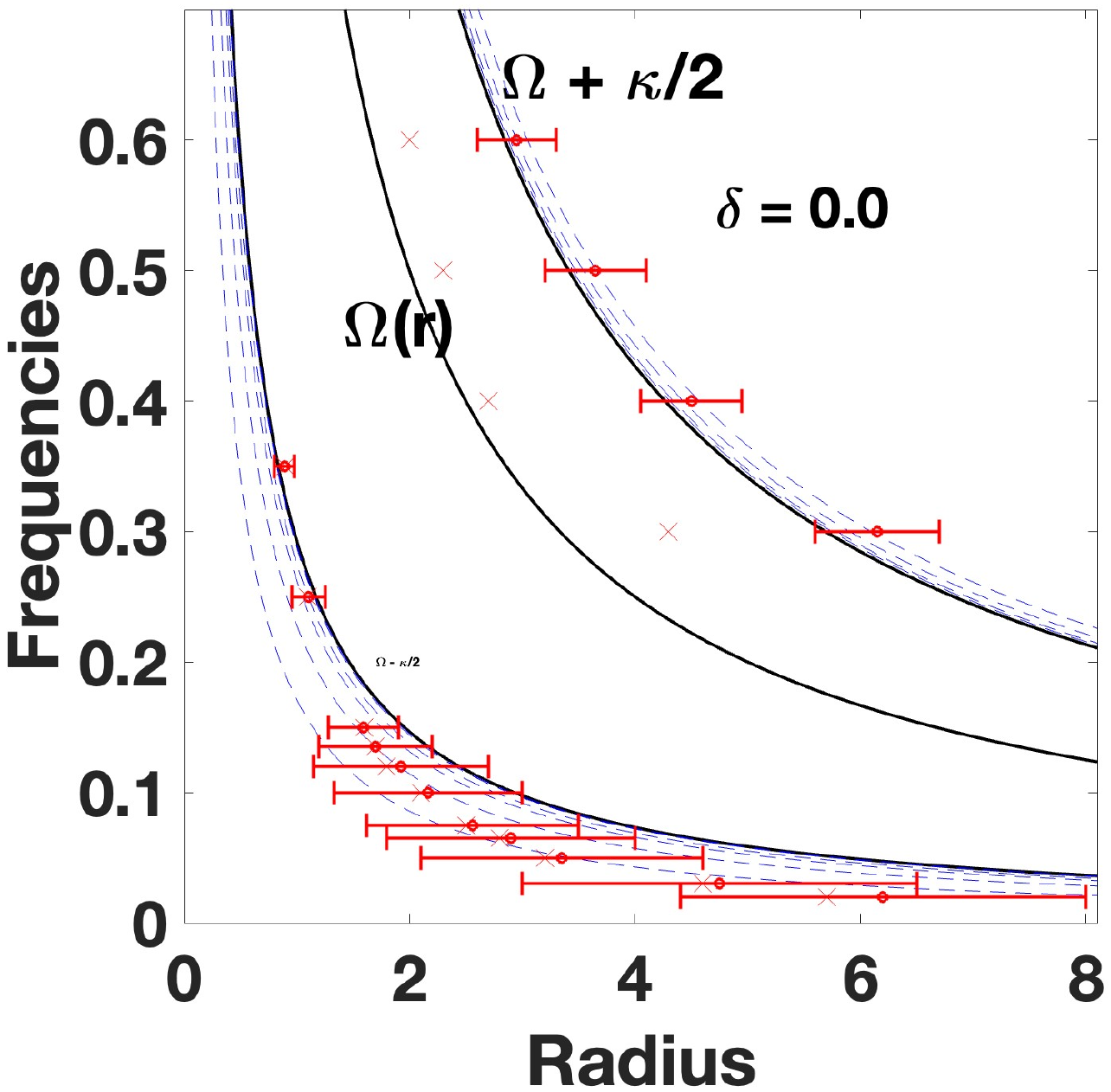}}
\caption{Like Fig. 3, except for the flat rotation curve potential, specified by the power-law index $\delta = 0.0$. Note also the addition of the upper solid curve for the $\Omega + \kappa/2$ locus, and dashed curves approximating the corresponding eccentric curves. The latter all lie close to the $(\Omega \pm \kappa/2)$ curves. The inner and outer radii for a number of Maximally eccentric stable REOs are shown again as horizontal error bars at a number of frequencies. Still more eccentric orbits at each frequency are sticky or unstable. Note that only near circular REOs are stable at the highest frequencies.} 
\label{fig:fig4}
\end{figure}

Again we find that the most eccentric REOs have very low pattern frequencies. Here too, based on their $\mathcal{R}$ values, bars comparable in size to these orbits would be classed as `slow.' And again, there are relatively low eccentricity REOs that are outer Lindblad resonance curves; several are shown in Fig. 4. It would be easy to nest these particular curves without crossings to form the backbone of a spiral or outer bar.

The most eccentric REOs shown in Fig. 4 are much more closely associated with the dashed, analytic resonance curves than in Fig. 3. They extend across several of these curves, so it is not possible to associate any particular REO with a specific curve, but there is better qualitative agreement than in the exponential/point-mass potential. This is not surprising since the analytic theory was derived for power-law potentials (see \citealt{struck06}). 

\subsection{Example 5: Falling and rising power-law rotation curve cases}

Fig. 5 shows frequency-radius diagrams for falling and rising rotation curve cases with indices $\delta = 0.3, -0.8$. (For reference, a Keplerian rotation curve has index $\delta = 0.5$, a solid-body case has $\delta = -1.0$.) The general trends in the frequencies of the maximally eccentric error-bars are similar to the previous cases. 

\begin{figure}
\centerline{
\includegraphics[scale=0.63]{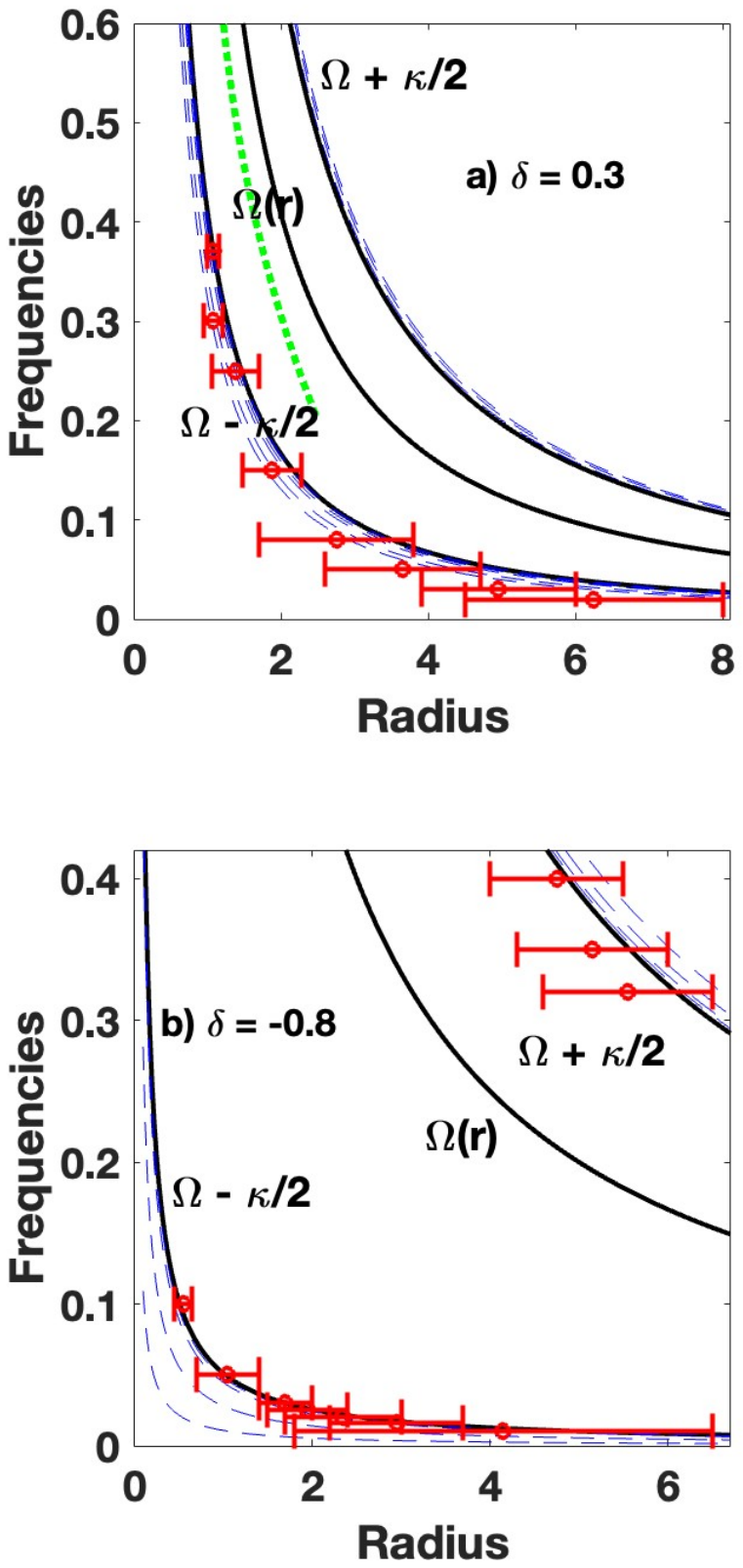}}
\caption{Same as Fig. 4, except with falling and rising rotation curve potentials, specified by the power-law indices $\delta = 0.3, -0.8$, respectively. The dashed green curve segment represents the pattern speed of a set of orbits excited by an ad hoc disturbance over a range of frequencies, representing a kinematic spiral wave. See text for details. } 
\label{fig:fig5}
\end{figure}

There are, however, some important differences, e.g., the $\Omega \pm \kappa/2$ curves are closer together in the falling rotation curve case of Fig. 5a than in Fig. 4. They are farther apart in Fig. 5b than in either of the other cases. This means that the $\mathcal{R}$ values of the maximally eccentric REOs in the Fig. 5a are relatively less than in the other cases, so the bars would be described as relatively fast. In the top panel of Fig. 13 below, where the large radius end of the error bars (for somewhat unstable or sticky orbits) are taken as the bar radius, we obtain $\mathcal{R}$ values less than the critical value of 1.4. Thus, bars based on these orbits would be classified as `fast' bars. But generally, those in Fig. 4 (see second panel in Fig. 13) and certainly those of Fig. 5b, would be the backbones of slow bars.  

 This trend of bar speed with potential form may generalize to concentrated versus diffuse potentials, i.e., the latter may generally host slow bars. This must also be reconciled with the evidence from N-body simulations that the buildup of central concentrations ends the growth of bars, see reviews of \citet{athanassoula13, sellwood14}. There are few examples in the literature with sufficient data to test these ideas against. But in accord with these ideas, the recent paper of \citet{buttitta23} shows that NGC 4277 has a slow bar embedded in the rising part of the rotation curve (diffuse), while NGC 4262 has a fast bar that extends out to the beginning of the declining (concentrated) part of the rotation curve. 

Another interesting feature of Fig. 5a is that the size of the maximally eccentric REO seems to vary less with frequency than in the flat rotation curve case of Fig. 4 or the case of Fig. 3. The opposite is true of the REOs in Fig. 5b, and those associated with an ILR only have significant eccentricity at very low frequencies. In the former case it would be easier to embed the smaller orbits within the larger ones over a significant range of frequency, without orbit crossings. This would, in turn, allow such orbits, with successive azimuthal twists a la \citet{lyndenbell72, kalnajs73} to make up the backbone of $n_r = 2$ spiral waves. In the case of Fig. 5b, this is possible in the vicinity of  an OLR. The $\delta = 0.3$ has no stable REOs with significant eccentricities associated with an OLR, so there could only be very tightly wound spirals based on near circular orbits. 

 \subsection{Section summary}

 In this section we have explored under what circumstances stable REOs, which might support bars, are found. Stable REOs associated with an ILR exist at a wide range of frequencies and over finite ranges of eccentricity in a variety of potentials. In most cases the maximum eccentricity for stable orbits is well below 1.0; the minimum is always 0.0. 

 In all the cases studied here there is a maximum frequency above which there are no stable REOs with non-negligible eccentricity. The value of this upper limit, in dimensionless units, varies between different potentials, see Figs. 3-5. 

 At frequencies where there are a range of REOs, and at eccentricities somewhat higher than the maximum for stable REOs, there generally exist quasi-stable orbits, which have the appearance of a thick REO. At still higher eccentricities orbits become 'sticky', i.e., appear to be stable for at least several circuits, but ultimately precess away. The stickiness decreases with increasing eccentricity, until the orbits become completely unstable at some eccentricity less than 1.0. This limit is discussed further below.

\section{Precessional evolution of REO ensembles for bar and spiral backbones}

Potential sources of disc perturbation and REO excitation include: low mass satellites colliding with the disc, massive companion flybys and other tidal interactions, internal formation of massive clumps (especially in young discs), and low level harrassment in dense group or cluster enviroments. For example, consider the idealized case of a tidal, impulsive perturbation on disc orbits that are all initially circular. Analytic studies of such cases have been provided by \citet{gerber94, donghia10}. While generically there is an impulsive change to both radial and an azimuthal velocity components, the latter is of more interest here. The azimuthal impulses induce changes in angular momentum, resulting in more eccentric orbits, and a better chance to excite REOs. 

We can write the azimuthal impulse as, $\Delta v_{\phi} = \Delta(R \Omega) \approx R \Delta \Omega$. The primary effect of an azimuthal impulse is to change the angular velocity. In radius-frequency (r-f) plots such as Figs. 3-5 this frequency change yields a translation up or down from the original position of a star; we will consider a downward, angular momentum losing case. With the assumption of initially circular orbits, the star will initially be on the circular frequency curve in the r-f plot. After the impulsive translation in the r-f plot, there are a number of possibilities for the structure of the perturbed orbit. The first is that the downward translation in the r-f plot is small and does not come near the $(\Omega - \kappa/2)$ curve. This leaves the star in a region without $n_r = 2$ REOs. On the other hand, if the star is pushed down close to the $(\Omega - \kappa/2)$ curve it can find itself on a REO.  

In this section we will explore some semi-quantitative examples to illustrate the range of possibility for REO excitation and wave backbones. Specifically, the p-ellipse approximation will be used to illustrate the kinematic evolution of ensembles of REO orbits chosen to illustrate the effects of particular types of disturbance. Only moderate eccentricity orbits, where the approximation is valid, will be considered. The use of analytic orbits makes it easy to include many orbits to illustrate complex waves. 

\subsection{Caustic spirals from REOs in different potentials}

We begin with examples of kinematic caustic waves, which are defined by an orbit crossing zone, and which can be produced by strong disturbances of discs, e.g., in galaxy collisions \citep{struck90, struck99}. The simplest are the ring waves in direct galaxy collisions, which are bounded by two fold caustics \citep{strucklotan90, struck10}. Spiral waves and tidal tails bounded by fold caustics can also be produced \citep{struck90, struck11}. These waves and their caustic edges are transient, though they may be repetitive until phase mixed. In impulsive interactions, the orbits that make up the waves are usually excited over a wide range of locations, with a corresponding range of precession frequencies, possibly including an ensemble of REOs. 

\begin{figure}
\centerline{
\includegraphics[scale=0.45]{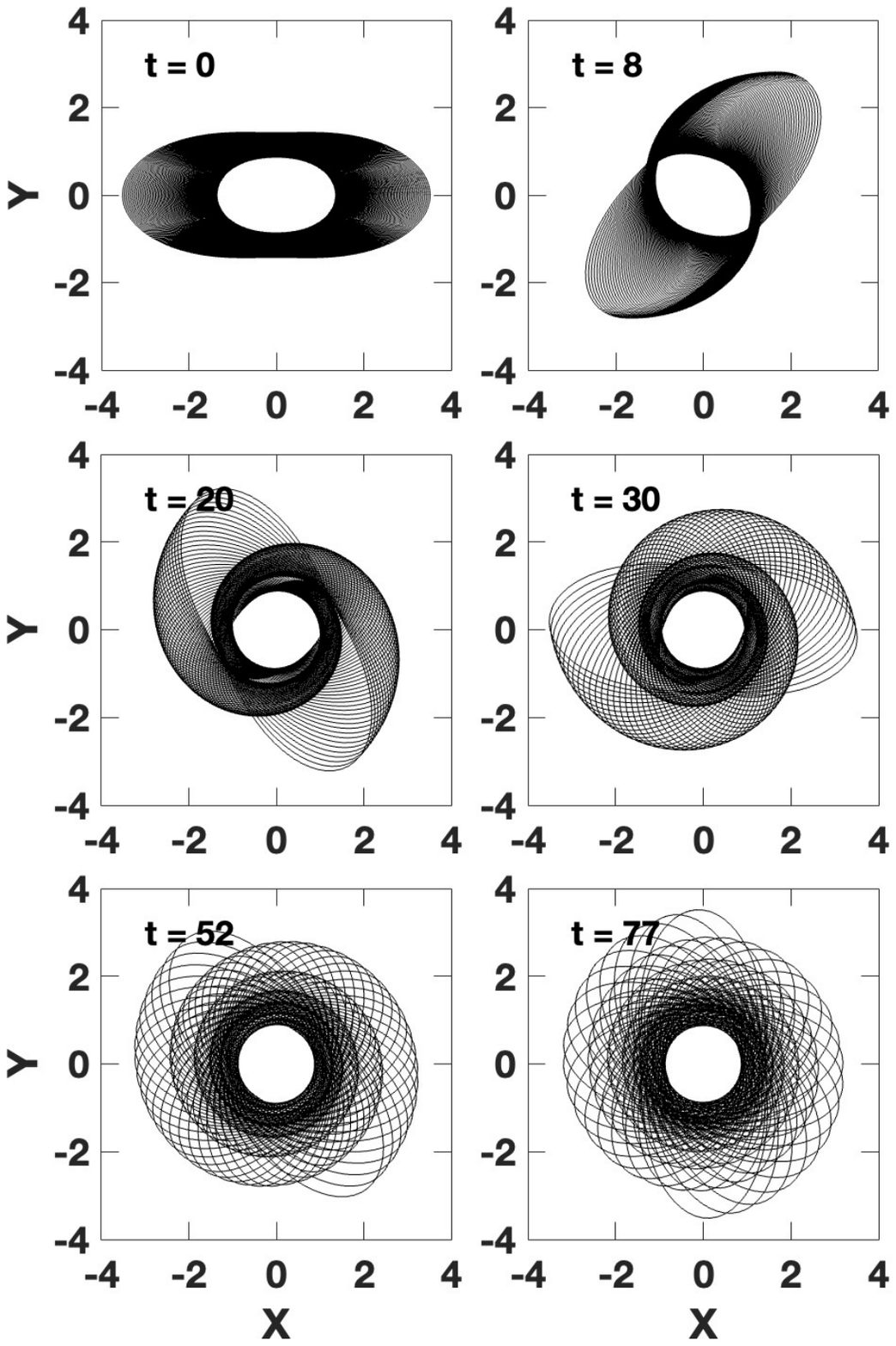}}
\caption{One hundred analytic approximate REOs, of submaximal eccentricity, were assumed to be excited along a nearly vertical line in the radius-frequency plane (like the green curve in Fig. 5)) in the potential with $\delta = 0.3$. Each subsequent panel shows a different dimensionless timestep of the precessional evolution.} 
\label{fig:fig7}
\end{figure}

\begin{figure}
\centerline{
\includegraphics[scale=0.45]{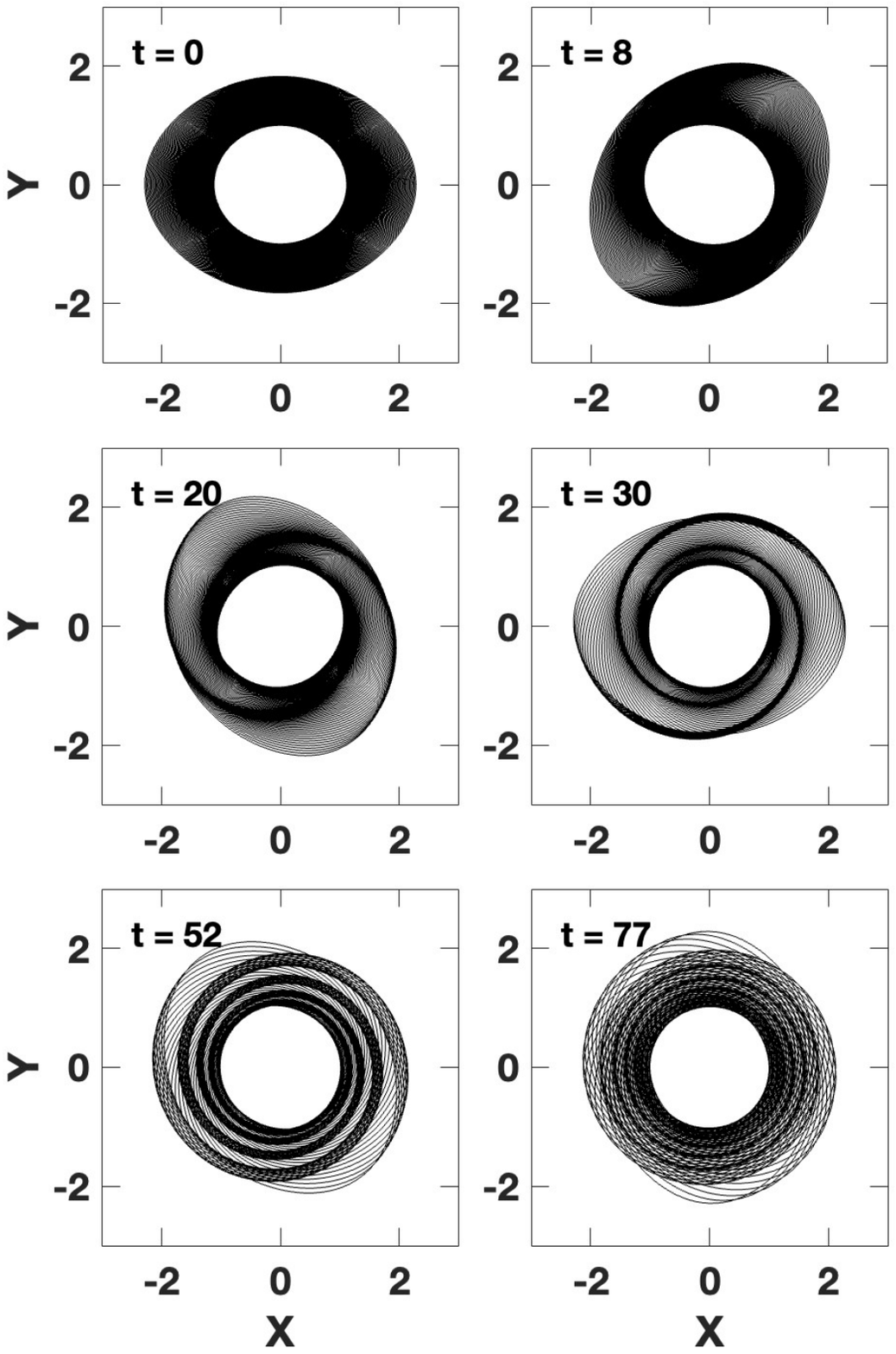}}
\caption{Same as Fig.~\ref{fig:fig7}, but in the slowly rising rotation curve potential ($\delta = -0.3$).} 
\label{fig:fig8}
\end{figure}

Figure~\ref{fig:fig7} shows a case where all the orbits derive from an ad hoc curve, illustrated in green in Fig. 5a, and assumed to be excited by some external disturbance. To make Fig.~\ref{fig:fig7} (and similarly for Figs.~\ref{fig:fig8}-\ref{fig:fig12}) 100 REO orbits at different radii along this line were initialized along a common semi-major axis, see the first panel of the figure. Like Fig. 5a, the potential is a $\delta = 0.3$ power-law (declining rotation curve). It was also assumed that the stars along the whole length of each REO orbit were excited. Specifically, points at 130 evenly spaced azimuths were chosen along each analytic p-ellipse in its individual pattern (precession) frame. These orbits were then evolved by computing their cummulative precession relative to the reference orbit. The orbital curves in the figures are drawn by connecting neighboring points on each p-ellipse by line segments at all times.  

The time series of panels in Fig.~\ref{fig:fig7} shows the development of kinematic spirals as caustic orbit intersections (e.g., at times of 8 and 20 units), and their subsequent widening and winding up. In this case, the widening of the orbit overlap region appears more important than wind-up. In consequence, the spirals disappear more by smearing due to radial phase mixing than by azimuthal stretching and wind-up. The final panel shows a rather shell-like appearance. In reality, this would be smoothed by scattering of the stars, and the cumulative effect of their random velocities. 

Fig.~\ref{fig:fig8} is produced in exactly the same way as Fig.~\ref{fig:fig7} except in a moderately rising rotation curve potential with $\delta = -0.3$. The appearance of the spiral waves is very different. They are more tightly wound from the time when they first become visible. They do not widen like the previous case, and they disappear more as a result of wind-up than of radial phase mixing. This case resembles the classical, Lin-Shu picture of tightly wound spirals than the previous case. However, in both cases of Figs.~\ref{fig:fig7} and ~\ref{fig:fig8}, there is no self-gravity. The case shown in Fig.~\ref{fig:fig8} is much like the kinematic approximation to a tidal spiral in Figure 13 of \citet{struck11}.

\subsection{Robust kinematic bars and associated spirals}

\begin{figure}
\centerline{
\includegraphics[scale=0.45]{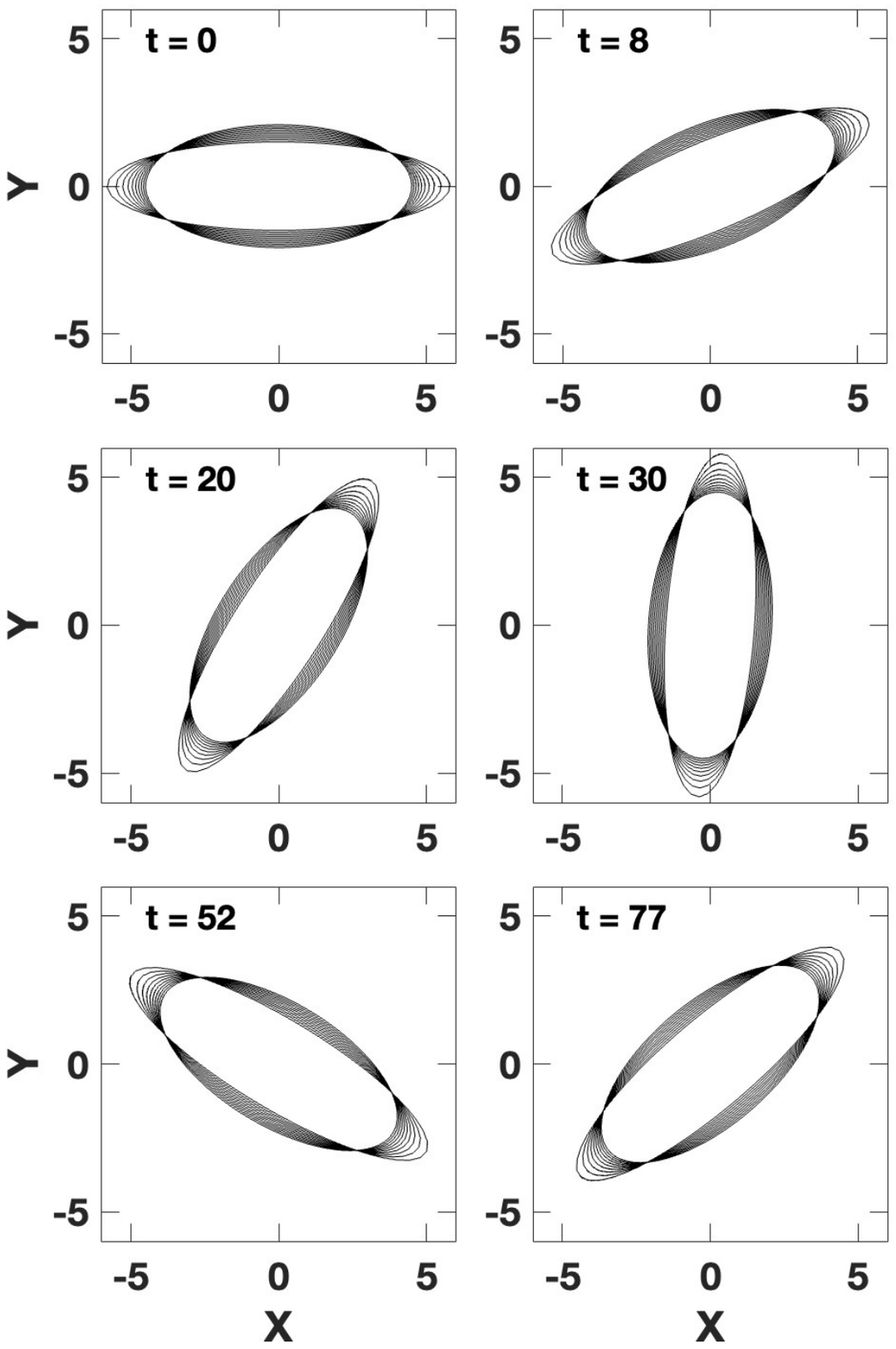}}
\caption{Snapshots of 10 analytic approximate, resonant orbits of slightly varying eccentricity chosen from a constant frequency line, and over a limited radial range in the $\delta = 0.0$, flat rotation curve potential. This illustrates a hollow backbone for a possible bar in a non-rising rotation curve potential.} 
\label{fig:fig9}
\end{figure}

Fig.~\ref{fig:fig9} shows the evolution of a bar formed of REOs excited at constant frequency, and over a small range of sizes and eccentricities, in a flat rotation curve potential ($\delta = 0.0$). The excitation of these orbits might be regarded as version of the Lynden-Bell theory of bar formation via the excitation of an ILR orbit, but with substantial eccentricity. These orbits are not strictly nested as in e.g., \citet{athanassoula92}, but rather cross at four points due to the modest eccentricity variations. If they were gas clouds, they would probably be dissipatively smoothed. Since the orbits converge with moderate relative velocities, we would not expect strong shocks at these points. As a kinematic feature made of superposed orbits this bar does not evolve. In reality, the orbit density may be great enough in the compressed regions to possess significant self-gravity. If so, it could also attract near resonant orbits in the manner envisioned by Lynden-Bell and grow into a stronger self-gravitating bar. 

This `hollow' bar orbits slowly in an absolute sense, with a frequency of $\Omega = 0.05$. The orbits are similar to that of one of the wider error bars of Fig. 4, which shows that it would also be viewed as having a fairly slow rotation in terms of the $\mathcal{R}$ parameter. We could construct a similar bar that would be classed as a fast $\mathcal{R}$ bar by choosing a similar range of REOs around an orbit like that of one of the wider error bars in Fig. 5 (i.e., in a declining rotation curve potential). 

\begin{figure}
\centerline{
\includegraphics[scale=0.45]{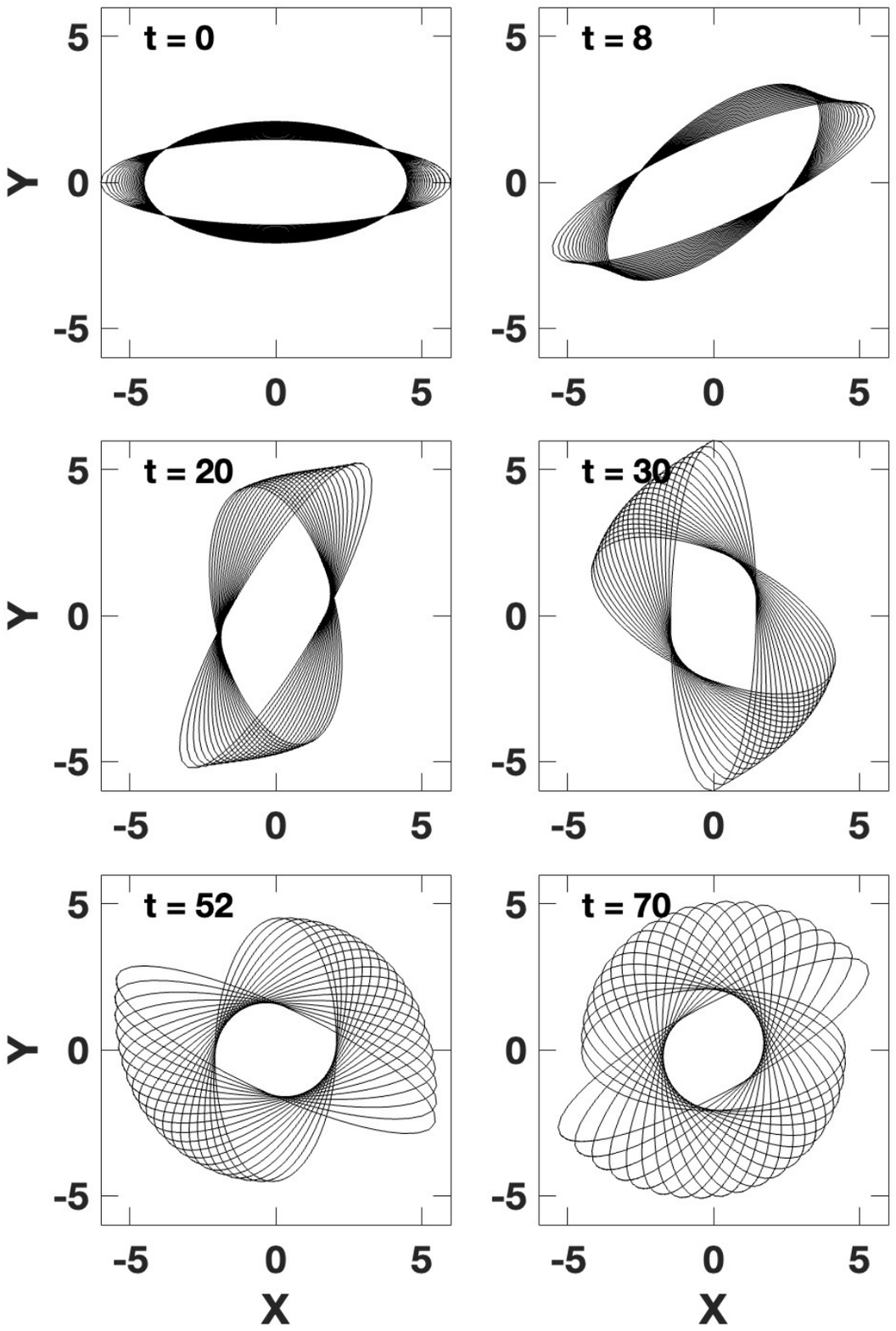}}
\caption{Same as Fig.~\ref{fig:fig9}, except the orbits all have slightly different precession frequencies (with a dimensionless frequency difference of 0.002 between each orbit), allowing the early bar-like morphology to become paired spirals, over a limited radial range, at later times.} 
\label{fig:fig10}
\end{figure}

Fig.~\ref{fig:fig10} shows a very similar case to that of Fig.~\ref{fig:fig9}, except instead of a strictly constant frequency, the initial orbits have a weak frequency dependence on size or eccentricity. Thus, after a short time the kinematic bar begins to distort. After sufficient time it begins to develop short, tightly wound, caustic spirals. By the last timestep there is little evidence of the bar. Because of the small orbital frequency range these spirals develop slowly and persist for a long time. 

With the appropriate frequency variation the spirals can develop on the outside of the bar rather than the inside. And in some such cases a small frequency variation in the inner orbits allows the bar to persist, giving a combined bar plus spiral structure.

\begin{figure}
\centerline{
\includegraphics[scale=0.45]{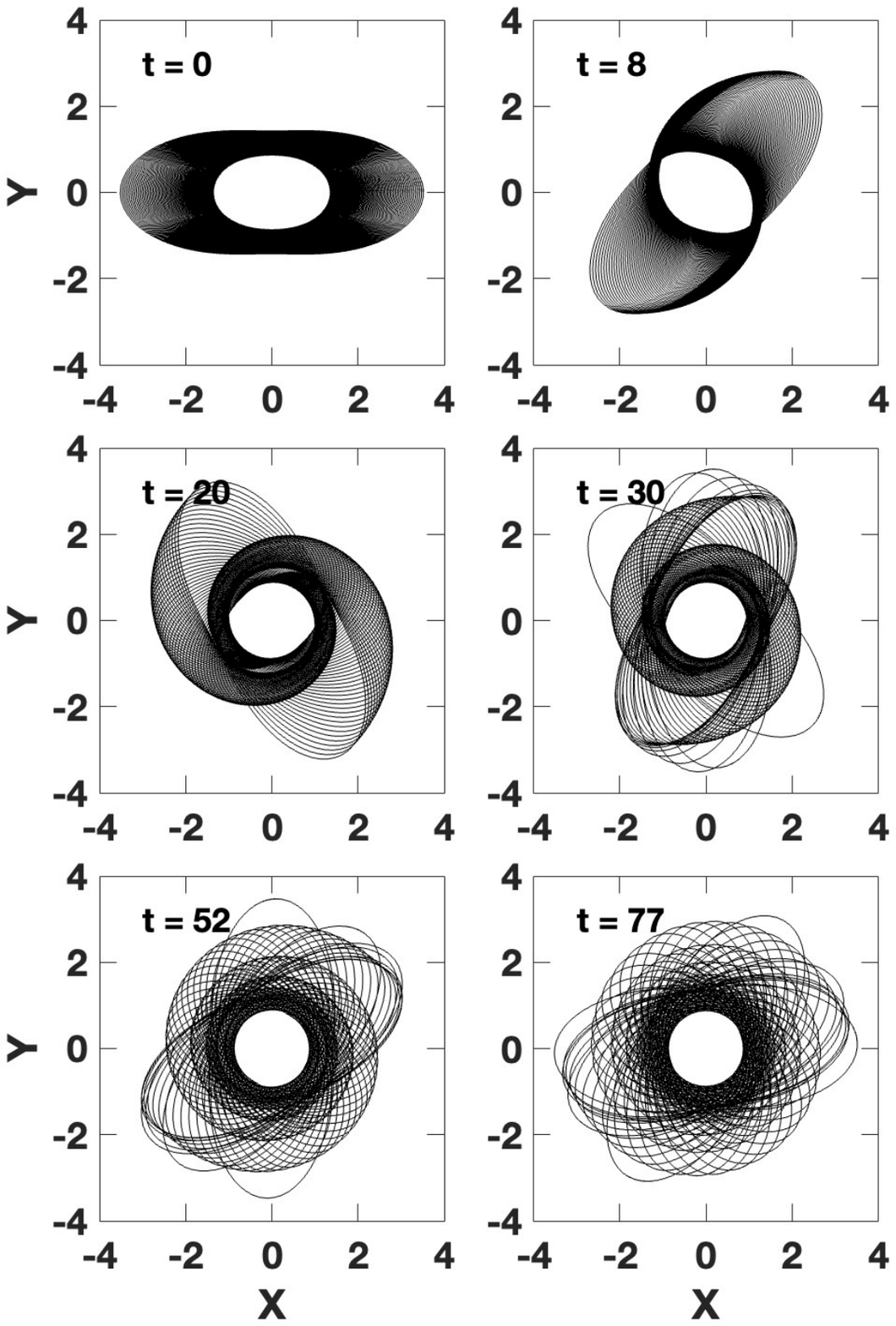}}
\caption{Same as Fig.~\ref{fig:fig7} (100 orbits in the $\delta = 0.3$ potential), but with a frequency increment of 0.0025 between each orbit, and a radius dependent initial angular twist between the orbits. See text for details.} 
\label{fig:fig11}
\end{figure}

Fig.~\ref{fig:fig11} shows a cases with a stronger frequency dependence, and with a declining rotation curve potential (a power-law with $\delta = 0.3$). This is the same case shown in Fig.~\ref{fig:fig7}, except each orbit is given an initial rotation by an angle which varies with orbit size. The initial twist is another perturbation parameter, in addition to frequency and amplitude. In this case an open and strong spiral pattern forms quickly, as in Fig.~\ref{fig:fig7}. This pattern also ultimately winds up and phase mixes. However, at the last times shown, a small set of orbits, remain synchronized for some time, and apparently could present a weak bar pattern. Thus, in this remarkable case, the kinematic waves could appear as a bar, a strong two-armed spiral, and then as a mostly phase mixed disc with a weak bar.

\subsection{Local resonant orbit excitation}

\begin{figure}
\centerline{
\includegraphics[scale=0.45]{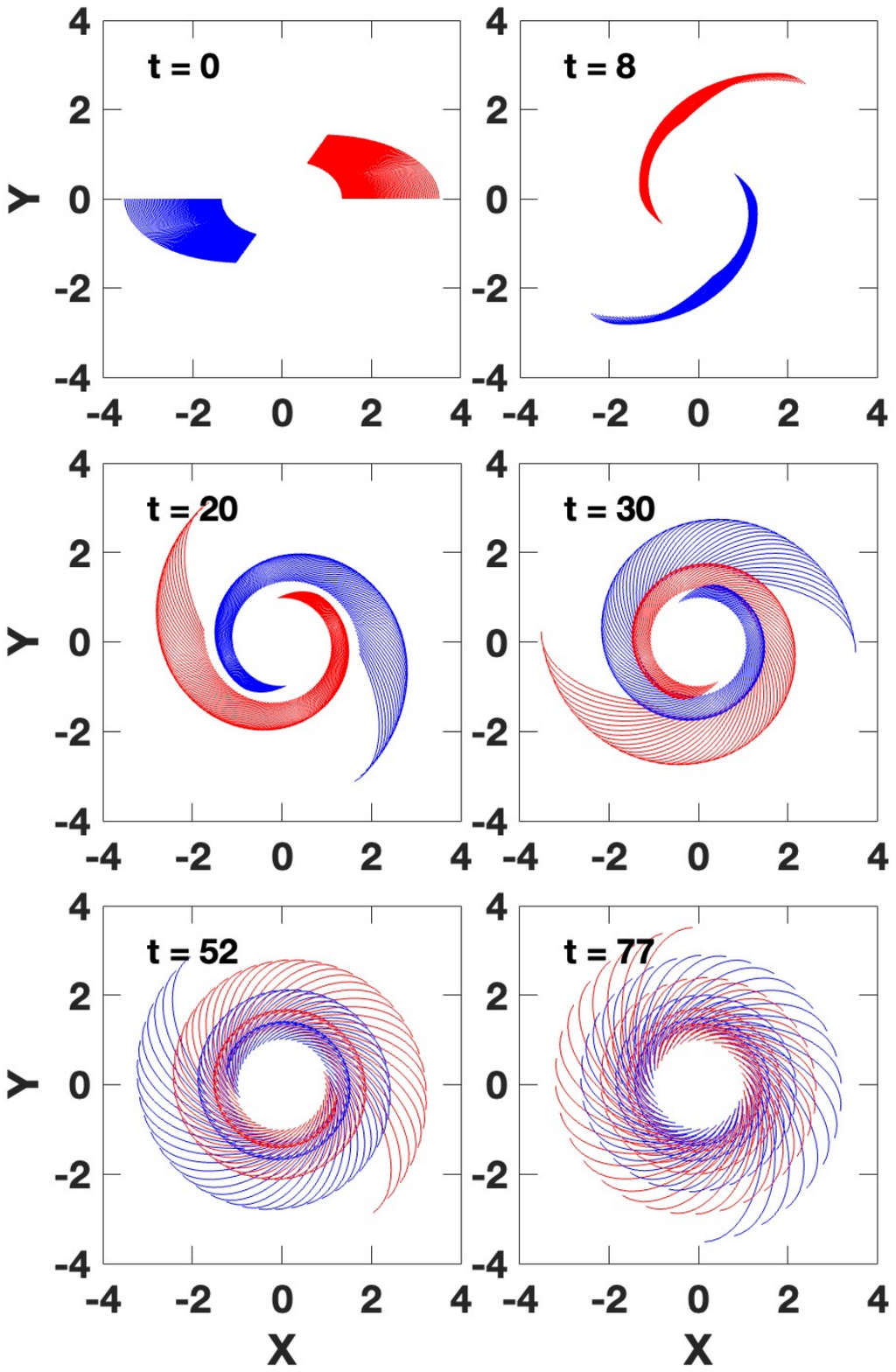}}
\caption{Same as Fig.~\ref{fig:fig11}, except particles are initially placed in a limited (but bisymmetric) azimuthal range to represent a localized disturbance, and its symmetric counterpart.} 
\label{fig:fig12}
\end{figure}

Up to this point we have considered waves produced by the excitation of REOs that have stars uniformly distributed around the whole orbit. For many perturbations in galaxy discs this is not very realistic. Fig.~\ref{fig:fig12} shows a case where stars are perturbed onto REO orbits over a limited range of frequency, size and azimuth, specifically in two patches. Each patch may represent the perturbation by a small satellite passing through the disc. Alternately, the two patches may represent the areas of strongest perturbation by an external tidal encounter. They could also represent clumps formed by local gravitational instabilities. 

We see that the particles quickly shear into dual spirals, and these subsequently wind-up and phase mix. The main point here is that as long as a significant range of frequencies are excited, shear will quickly distribute the stars in azimuth, and form waves like those seen in previous cases. The exceptions are in solid body rotation curves, and cases of single frequency excitation like that shown in Fig.~\ref{fig:fig9}. In such cases, the initial patches in Fig.~\ref{fig:fig12} could persist as clumps.

\subsection{Section summary}

The previous section described the range in frequency and eccentricity of REOs, while this section focused on kinematic wave backbones that can be made from ensembles of these orbits. We began with a few comments about how REOs can be excited, and a definition of kinematic caustic spirals. Then, examples were given of the variety of possible bars and spirals based on REO backbones. The section concluded with an example of spirals resulting from very localized REO excitations.

{\it It is clear from even these few examples that depending on the various parameters, including potential form, range of excitation frequency and amplitude, and variations in excitation azimuth and time as a function of position, a huge variety of outcomes are possible. Some of these are very transient, some are persistent.} We will explore some of the implications of this result for observed waves and their systematics in the next section.

\section{Discussion 1: Ramifications to the theory of bars}

The theory of waves supported by REO backbones, excited in the extended resonance regions around major Lindblad resonances, links to and overlaps several theories and observations of bars and spirals. In this and the following two sections we will consider some of these links.

\subsection{REOs and the Lynden-Bell process}

\citet{lyndenbell79} suggested that the gravity of an eccentric orbit at the ILR and other near resonant orbits in a nascent bar could capture other nearby orbits into oscillations around the bar, and thus, progressively build and strengthen the bar. This idea has been widely accepted and studied in the epicyclic approximation, but has some potential problems. The first is that orbits very near the ILR, e.g., approximated by small epicycles, are wide ovals rather than narrow bars. Orbits for more eccentric REOs are centred at modest offsets from the ILR as shown in Fig. 3-5. The existence of these REOs allows for narrower bar backbones, although the instability of the most eccentric REOs may still be a problem.

A second difficulty in the epicyclic theory is that it may be difficult to build a bar from an isolated ILR orbit and a small population of near resonant orbits. It could take a good deal of time to build up substantial gravity in this nascent bar and the orbital coherence could be disrupted in the meantime. N-body simulations generally show rapid bar formation. A common solution to this problem is to assume that a radial region with a near harmonic potential supports a range of nested REOs (see the reviews of \citet{athanassoula13, sellwood14}).  

On the other hand, examples like that shown in Fig.~\ref{fig:fig9} illustrate how multiple REOs excited at a single or small range of frequencies could immediately provide a larger population of resonant and near resonant orbits, presumably with a larger gravity in the feature. This would facilitate a much more rapid buildup in a young bar, which could be quite narrow from the outset, in a variety of potentials.

\subsection{Bar lengths, eccentricities, and speeds}

It seems to be well established in the literature that most bars are `fast' in the sense that these bars have values of the speed parameter (corotation radius to maximum bar radius) close to $\mathcal{R} = 1.4$ \citep{aguerri03, rautiainen08, corsini11, aguerri15, cuomo19, cuomo20}. There are some slow bars, however, e.g., \citet{buttitta22, cuomo22}. These may be more common in late-type galaxies \citep{rautiainen08, williams21}. There also may be ultrafast bars with lesser values of this parameter, but some of these may be the result of inaccurate bar length estimates \citep{cuomo21, garma22, peterson23}. 

In Sec. 3.2, we considered limits on the eccentricities of REO orbits, as a function of orbital frequency and potential form. Beyond these limits the orbits generally become 'sticky' (see \citealt{athanassoula13}), and eventually evolve away from the REO form. See the right hand panels of Fig.~\ref{fig:fig13} below. The maximum length of a stable REO in a given potential at a given frequency implies a maximum eccentricity of its backbone orbits, so this maximal orbit may determine both the length and shape of the bar. Consider the two panels on the left of Fig.~\ref{fig:fig13}. These panels are like those in Figs. 4 and 5a with the red error bars representing the maximum size of nearly periodic orbits. In both of the potentials shown (and the $\delta = -0.8$ one in Fig. 5) the radii of the maximally eccentric REOs do not reach much past the $(\Omega - \kappa/2)$ curve. Bars of length about equal to these maximal radii would be classed as slow.

On the other hand, sticky or wobbly orbits can persist within the pattern for many orbits, as seen in the panels on the right side of Fig.~\ref{fig:fig13}. These right hand panels show a series of orbits (all in a common pattern frame) in the $\delta = 0.3$ potential, with increasing size and eccentricity from top to bottom. The third and fourth panels from the top are sticky, i.e., they stay in the pattern for some time before beginning to precess away, as shown. The orbit of the fourth panel has a three-loop form. The fifth panel is similar except that it only stays in the pattern for a couple of orbits, and then precesses rapidly away. The sticky orbits can contribute to the backbone. So we can adopt them as a maximal measure of the bar length and shape in the outer bar. 

The long, cyan colored error bars in the left panels of Fig.~\ref{fig:fig13} give the extent of such maximal sticky orbits at different frequencies. In the upper left panel of Fig.~\ref{fig:fig13} we can see that bars with lengths equal to these orbits are fast, with values of about $\mathcal{R} = 1.4$. This is also true for the highest frequency orbits in the lower left panel, but not for lower frequency cases. These panels also show a couple of solid red curves; the uppermost in each panel is the $\Omega - \kappa/4$ curve. It appears that the fast orbits do not go to much larger radii than these curves in these potentials, see \citet{elmegreen96, michel06} for similar conclusions on bar lengths and the 4:1 resonance. Slow orbits do not reach them. In rising rotation curves (not shown in Fig.~\ref{fig:fig13}), the extremal orbits are all slow. 

The lower red curves Fig.~\ref{fig:fig13} also represent resonant locii. In the upper left panel, the lower curve is the $\Omega - 3\kappa/4$ curve. The $\Omega - 2\kappa/3$ curve is also near that curve, but not shown. In the lower left panel the $\Omega - 3\kappa/4$ curve  does not exist (is not positive), and the $\Omega - 2\kappa/3$ curve is shown instead. The corresponding resonances may also play a role in determining the minimum radius of the extremal, quasi-stable REOs. 

In many barred galaxies the bar likely extends beyond the rising rotation curve region and into a flatter or falling region. There extremal orbits like those in Fig.~\ref{fig:fig13} can be excited, and these may determine bar shape and length.  

\begin{figure}
\centerline{
\includegraphics[scale=0.42]{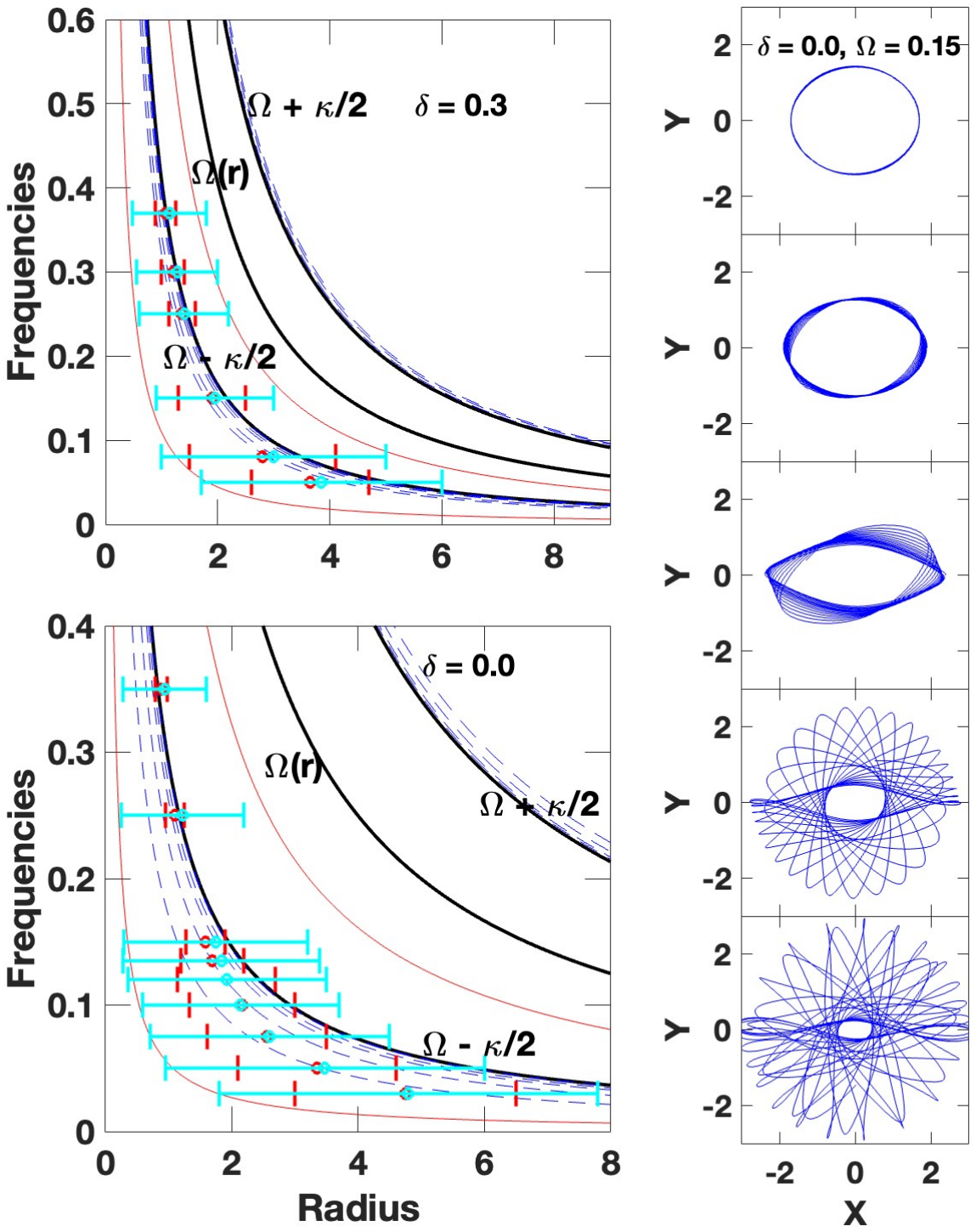}}
\caption{The left hand panels are like Figs.~\ref{fig:fig4} and \ref{fig:fig5}, with the extent of the maximally eccentric orbits shown by red error bars. Maximal transiently stable or `sticky' orbital extents are shown by cyan error bars. Black curves show the $\Omega - \kappa/2$, $\Omega(r)$ and $\Omega + \kappa/2$ frequencies vs. radius. The upper red curves in these panels show the $\Omega - \kappa/4$ curves, the lower red curve in the top left panel shows the $\Omega - 3\kappa/4$ curve, and the lower red curve in the bottom left panel shows the $\Omega - 2\kappa/3$ curve. The right hand panels show a sequence of stable to increasingly unstable orbits at a single frequency in the $\delta = 0.3$ falling rotation curve potential. See text for details.} 
\label{fig:fig13}
\end{figure}

\subsection{REOs may constrain bar evolutionary paths}

Simulations show that once bars become self-gravitating, they interact with other galaxy components like the spheroid and halo, and continuously lose angular momentum (see reviews of \citealt{athanassoula13, sellwood14}, for the theory of this process see \citealt{tremaine84, chiba23}). N-body models also show that as bars slow (in absolute terms) they become longer and stronger. (Indeed, a number of numerical simulations also show a slowing in terms of the $\mathcal{R}$ parameter, e.g., \citet{oneill03, algorry17}.) Direct observational support for the evolutionary slowing is hard to obtain, but see e.g., \citet{geron23}. A priori it is unclear to what degree the systematics of kinematic bars continue to hold for gravitating bars. The study of \citet{struck18} indicates that it is harder to maintain the full array of REO orbits in the presence of power-law bar potentials. However, this study was carried out at relatively high pattern speeds, where the results above suggest very eccentric REOs do not exist even in symmetric kinematic models.

None the less, let us assume that some REOs, and associated trapped orbits, still provide a backbone for bars during their evolution. Suppose, in particular, that the largest of these orbits maintain sizes comparable to the bar itself. Then as the bar slows these orbits can become still larger, since according to Figs. 3-5 and \ref{fig:fig13} above, lower frequency bars generally have longer and more eccentric maximal REOs. The increasing eccentricity of the backbone REOs would lead to thinner bars.

The assumption that the backbone REOs grow with the slowing bar, and that this growth in REO size and eccentricity drives the corresponding growth in the bar as a whole, is one possibility. The actual outcome depends on the transfer of angular momentum within the bar, and specifically, how the angular momentum of the largest backbone orbits change and how the most eccentric orbits are excited.

For example, consider the specific case where the change in the angular momentum of the maximal REO is negligible. Suppose further that we can approximate this evolving maximal REO with a p-ellipse form, though this may not be very accurate when the eccentricity becomes high. According to equation B1 of \citet{struck06} the specific angular momentum, $h$, of a p-ellipse orbit is given by,

\begin{multline}
\label{eq7}
\frac{h}{{(GM_*)}^{1/2}} = p^{1-\delta} \\
= 2a
\left[ \frac{(1-e^2)^{1/2+\delta}}{(1-e)^{1/2+\delta}
+ (1+e)^{1/2+\delta}} \right]^{1-\delta}
= \left( \frac{a}{g_o(e)} \right) ^{1-\delta},
\end{multline}

\noindent where $M_*$ a scale mass, and $a$ is the semi-major axis of the orbit. We can take the time derivatives of this equation, and assume that the derivative of the left-hand-side (i.e., of $h$) is zero, to obtain, 

\begin{equation}
\label{eq8}
a\frac{da}{dt} = \frac{1}{g_o} \frac{dg_o}{de}
\frac{de}{dt},
\end{equation}

and, 

\begin{equation}
\label{eq9}
\frac{a}{a_i} \simeq \frac{g_o}{g_{oi}}
= \frac{(1-e)^{1/2} + (1+e)^{1/2}}{2(1-e^2)^{1/2}}
\simeq \frac{1}{2(1-e)^{1/2}},
\end{equation}

\noindent where subscript $i$ denotes initial values. In the last two equalities we assume that $e_i$ is small, and so, $g_{oi} \simeq 1$, and the last equality assumes relatively large values of $e$. This equation shows that $e$ and $a$ grow (or decline) together when there is little change in angular momentum. This is consistent with a downward evolution to larger more eccentric orbits, with diminishing frequency in Figs. 3-5.

Alternately, consider the case where the evolution of the largest orbit follows a track of decreasing frequency through the maximally eccentric EROs. In all physically relevant power-law potentials (i.e., $-1 < \delta < 0.5$) the angular momentum of these orbits increases modestly with decreasing frequency. Since N-body models show that angular momentum is transferred outward in evolving bars, this case may be more realistic than the constant angular momentum case. It yields faster growth of a narrower bar than that case. 

\subsection{Effects of central concentrations and evolution on bar speeds}

Numerical models show that bars tend to weaken with the growth of central concentration \citep{norman96, das03, shen04, athanassoula05, bournaud05, debattista06}, or be difficult to form in the presence of such concentrations \citep{jang23}. It is clear from Fig. 3 (especially the top panel) that centrally diffuse potentials, e.g., with rising rotation curves, only allow moderately eccentric REOs at low orbital frequencies. The latter also have very small $\mathcal{R}$ speeds. Although REOs are not shown explicitly, this is also true of the inner regions of most of the potentials illustrated in Fig. 1. Centrally concentrated falling rotation curve potentials, as shown in Figs. 5 and \ref{fig:fig13}, allow quite eccentric orbits at higher frequencies, and higher $\mathcal{R}$ values. 

It is generally believed that bars are found in, and likely form in, the rising rotation curve part of the disc potentials. Yet, as noted above, most bars have fairly high $\mathcal{R}$ values, which are only supported by very eccentric REOs in flat or falling rotation curves (Fig.~\ref{fig:fig13}). This contradiction can be resolved by the bar evolution to slower speeds, but longer lengths. The latter evolution can take the maximal REOs out to radii where the rotation curves flattens or begins to decline (as in the middle right panel of Fig. 1). The excitation of these larger and flatter orbits would allow an increase in $\mathcal{R}$ values for a bar with these backbone orbits, while its pattern speed decreases. I.e., as the bar slows in frequency, it speeds up in $\mathcal{R}$.

\subsection{Bar surface density profiles}

\citet{elmegreen85, elmegreen11, elmegreen96, kim15} have found that bars commonly have one of two type types of surface density profile, either flat (in early type galaxies) or exponentially declining (in late types). The flat profile is easy to understand as simply inherited from the pre-existing profile of the rising (e.g., solid body) rotation curve region. That is, assuming the inner disc is dominated by its own gravity, not the dark halo. If bars grow beyond that region, and beyond an exponential scale length, into a flat or declining rotation curve, then we can understand the exponential profiles as also inherited. 

\citet{anderson22, beraldo23} have argued that bars can also form flat shoulders as they grow and strengthen. They further suggest that the shoulders are produced by the excitation of more $x_1$ loop orbits. This qualitatively agrees with the notion that as a bar slows it moves into regions with more eccentric maximal REOs. It would be interesting to know if the shoulders are confined to regions with stable REOs, or extend into the region of transiently stable orbits.

\subsection{Section summary}

The existence of a range of REOs at a variety of frequencies in many galaxy potential types makes it easier for any perturbation to excite resonant orbits than if the only such orbits were near circular with frequencies very near $\Omega - \kappa/2$. This in turn makes it more likely that the LBK process plays a large role in generating or maintaining bars. At any given pattern frequency, if bar lengths, eccentricities and speeds relate directly to supporting REOs, then these quantities will be constrained by the maximum stable REO at that frequency. Similarly, if angular momentum transfer to REOs is efficient in evolving bars, then their evolving structure will be constrained by the sequence of maximal REOs as a function of frequency. Specifically, bars with decreasing pattern frequencies, may grow from rising rotation curve regions into flat rotation curve region, and go from fast to slow $\mathcal{R}$ values. Newly excited $x_1$ orbits in this process may add shoulders to the bar profile.

\section{Discussion 2: Kinematic, caustic spirals}

In Section 4 we saw that a variety of spiral waves could be produced by the overlap of REOs. In this section we will see that these waves can help understand a number of the systematic properties of spiral waves described in the literature. 

\subsection{Wind-up and recurrence}

Figures~\ref{fig:fig7} and \ref{fig:fig8} show examples of kinematic spirals in the spirit of the famous example of \citet{kalnajs73}. As is well known and shown in the figures these spirals generally wind up or phase mix on short to intermediate timescales, depending in part on the potential. The orbits in these figures wind up because they are drawn from a significant range of precession frequencies. These waves could represent common transients in galaxy discs, and may be especially important in galaxy interactions \citep{struck90, struck12}. 

The orbits in Figs.~\ref{fig:fig7} and \ref{fig:fig8} consist of p-ellipse approximations, which do not take into account the result that at a given pattern frequency, stable orbits only exist up to a specific maximum eccentricity. Thus, though they could persist for a number of orbital periods, some of the orbits in Figs.~\ref{fig:fig7} and \ref{fig:fig8} may be only metastable. {\it Kinematic spirals will only be long-lived in potentials where a wide range of eccentricities are stable in significant ranges of frequencies.} 

\begin{figure}
\centerline{
\includegraphics[scale=0.39]{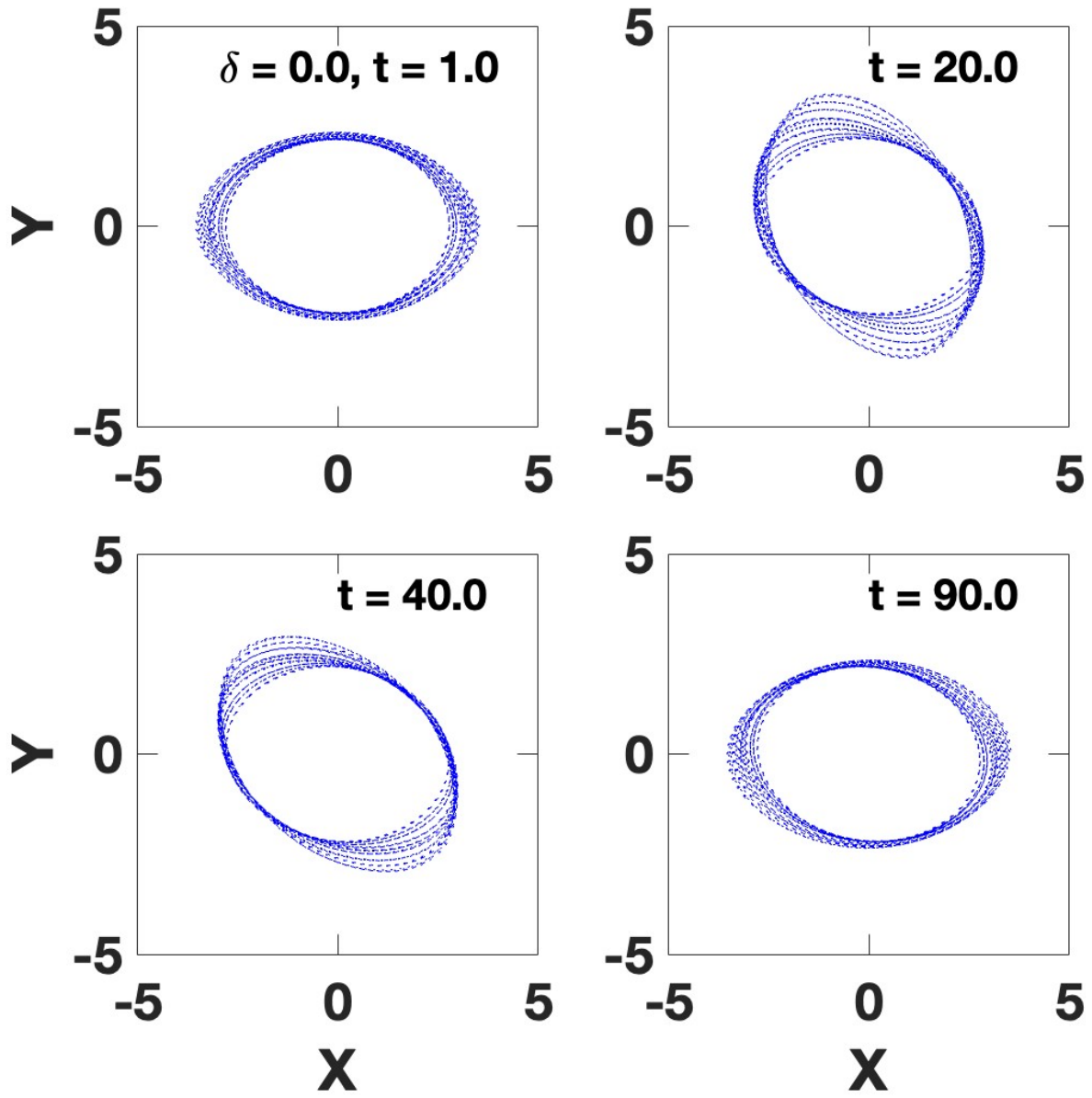}}
\caption{Nine orbits computed in the flat rotation curve potential are shown. These orbits have frequencies that differ by small amounts: 0.075, 0.073, 0.071, 0.069, 0.067, 0.065, 0.063, 0.061, 0.059 in dimensionless units. The orbits are aligned along the x-axis initially, then each is successively offset by $\pi/20$ radians from the smallest outward. The evolution of their differential precession in the frame of the smallest orbit is shown at different times in the successive panels.} 
\label{fig:fig15}
\end{figure}

Figures 3 and 4 show that at least in limited regions there can be nested orbits with a range of eccentricities, and which lie within a relatively small range of frequencies. Relatively long-lived spirals could form around ensembles of such orbits. Fig.~\ref{fig:fig15} provides an example, with 9 orbits in a flat rotation curve potential. It is assumed that particles at all initial azimuths around the orbit are excited. These orbits have small precession frequency differences, and small eccentricity differences to allow the nesting. The particular eccentricities are all similar, and chosen to be well within the stable range for the chosen frequencies in the given potential. These localized REOs might be excited, for example, via an interaction with a small satellite galaxy.

After the orbits have been given their small initial offsets, they precess due their different frequencies, and each panel in Fig.~\ref{fig:fig15} shows a different stage of this evolution, with times indicated on the figure. Panel a) of Fig.~\ref{fig:fig15} shows a time shortly after the start of this precession, when the orbits are still nested. At the time of panel b) orbital overlap has formed kinematic spirals, and a region of lower orbital density between them. The visibility of the orbital overlap spirals could be enhanced by gas pileups and star formation. 

Panel c) of Fig.~\ref{fig:fig15} shows that the kinematic spirals can persist for a fairly long time, i.e. for more the $40/2\pi$ orbital periods at a radius of 1.0 units. However, these spirals do eventually wind-up in their own fashion, and the final panel shows that they return to their original relative orientations. This cycle will repeat until the individual orbits are sufficiently disturbed to lose their coherence. 

\begin{figure}
\centerline{
\includegraphics[scale=0.39]{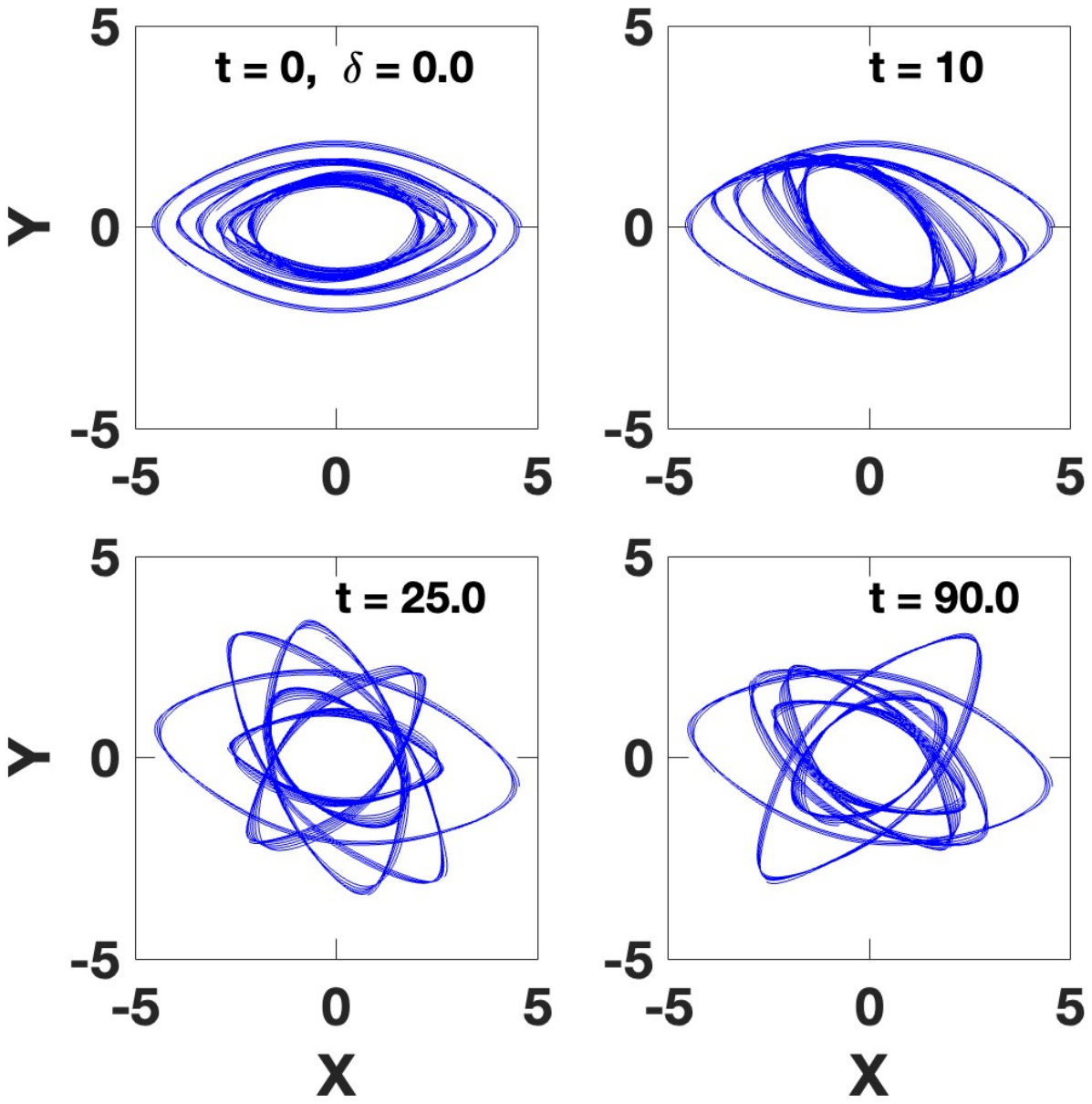}}
\caption{Six orbits computed in the flat rotation curve potential with frequencies that differ by larger amounts than in Fig.~\ref{fig:fig15}: 0.050, 0.065, 0.075, 1.00, 1.20, 1.35 in dimensionless units. The orbits are aligned along the x-axis initially, but not successively offset in angle as in Fig.~\ref{fig:fig15}. The orbits chosen are larger and more eccentric than those in Fig.~\ref{fig:fig15}. Each is near the stability boundary, and so, the superposition of several orbits has significant width. The evolution of their differential precession in the frame of the large orbit, in this case, is shown at different times in the successive panels.} 
\label{fig:fig16}
\end{figure}

The four snapshots in Fig.~\ref{fig:fig16} show a quite different case, though again in the flat rotation curve potential. The figure shows the differential precession of six near maximally eccentric REOs with a larger range of frequencies than in Fig.~\ref{fig:fig15}. By the time shown in panel b) there are very open spiral arms, as in Fig.~\ref{fig:fig11}. As shown in panel c), these structures do not wind up so much as they unravel. Aspects of this kind of evolution can also be seen in the later panels of Fig.~\ref{fig:fig11}. This evolution is not periodic, at least on any reasonable timescale, in contrast to the case shown in Fig.~\ref{fig:fig15}. However, as evident in panel d) we do see the sporadic occurrence of transient, small spirals made up of the overlap of two or three orbits. {\it Thus, the excitation of these kinds of REOs could drive the continuing formation of flocculent spirals.} Flocculent spirals are generally assumed to result from local gravitational instabilities, but both processes could operate simultaneously and synergistically.

\subsection{Spiral pattern speeds and pitch angles}

One of the most basic questions about spiral waves is how the pattern speeds of the waves vary with radius. In the classic version of Lin-Shu density wave theory it is assumed that the pattern is rigid \citep{lin64, lin66, bt08}. Simulation results often show prompt windup of induced spiral waves (e.g., review of \citealt{dobbs14}), a natural consequence of differential pattern speeds. On the other hand, some models have shown persistent or saturated spirals \citep{donghia13,saha16, sellwoodcarlberg21, sellwoodcarlberg22}. Observational evidence is mixed, with some evidence for near constant pattern speeds \citep{peterken19, meidt09}, and other evidence for radial variation \citep{merrifield06, speights11, speights12}. 

To provide the backbone for a spiral wave, we need an ensemble of nested, or nearly nested, REO orbits. The examples above also show that if we want significant overlap in these backbone orbits and to obtain a reasonably open spiral, then the orbits must have significant eccentricities. As can be seen in the sample potentials of Figures 3, 4, and 5, these joint constraints are very restrictive. Those figures make clear that it is generally not possible to satisfy those requirements in the central regions of centrally concentrated potentials, at high frequencies or pattern speeds.

On the contrary, at low pattern speeds those figures show that most potentials have regions of REO orbits which range up to quite high eccentricities. In many potentials, it is also possible to nest these over a modest range of frequencies. High eccentricities allow nearby orbits to overlap to a greater degree with substantial pitch angles. If the frequency range is small, then the windup of these orbits will be slow. As is apparent in the examples and figures above, spirals made of nested REO orbits extending only over a narrow range of frequencies are probably unusual. Those with REO orbits extending over a significant range of frequencies will have pattern speed variations with radius.  

In sum, wave backbones made up of nearly circular REO orbits will be tightly wound; those made of eccentric orbits will be much more open. The former orbits can exist at almost any frequency near an ILR (or an OLR). The latter tend not to exist near central regions, especially those with a high central concentration. They are more likely to be found at low frequency, well out in the disc, and somewhat offset from the $(\Omega - \kappa/2)$ curve in pattern speed (see Fig.~\ref{fig:fig13}). 

Another correlation found recently in observational studies is between spiral pitch angle and disc shear \citep{seigar05, seigar06, grand13, yu19}. In N-body models evolving pitch angles have been studied, and interpreted as the result of, and in accord with, swing amplifier theory \citep{baba13, michikoshi16, michikoshi18}. The connections between the observations and models have not been studied in detail, but see discussion in Sec.2.2.1 of \citet{dobbs14}.

The shear is a direct function of the gravitational potential, and is minimal (maximal) in rising (falling) rotation curve potentials. In fact, \citet{yu19} find a stronger correlation between pitch angle and rotation curve structure than with measured shear. Figures 4 and 5 show a trend in the maximal REOs near OLRs in accord with the shear-pitch angle correlation. For example, the rising rotation curve case ($\delta = -0.8)$) has quite eccentric REOs, that could be easily nested over a modest frequency range, making a good spiral backbone. \citet{roca13} found the spiral arms in their N-body models to be `nearly corotating with disc particles.' The most eccentric REOs in this case are not too distant from the corotation curve in Fig. 5, and since they would frequently be observed near apoapse, with an azimuthal velocity less than the local circular velocity, this appearance would be enhanced. The flat rotation curve case shown in Fig. 4 only allows smaller eccentricities in OLR REOs, and thus, more tightly wound spirals with the greater shear. The falling rotation curve case (e.g., the $\delta = 0.3$ case) only allows near circular REOs on the OLR, or very tightly wound spirals with relatively high shear. These trends are not as clear for REOs near an ILR, though they do seem to hold at large radii and low frequencies. Thus, if REO backbones are important in spirals the shear-pitch angle correlation can be understood as a result of the limits on maximally eccentric REOs in different potentials. 

\subsection{Section Summary}

Compressed and overlapping REOs can form the backbone of a wide variety of kinematic spiral waves, and they can have pattern speeds that are constant or vary with radius. The pitch angles of these waves depend primarily on the eccentricity of the constituent REOs. Since the range of allowed eccentricities depends on the potential form, or central concentration, this can help understand the sequence of opening angles along the Hubble sequence.

\section{Summary and Conclusions}

This paper has focused on the characteristics of families of REOs associated with the ILRs and OLRs of galaxy discs in two dimensions. The primary results for single component, monotonic potentials include the following. 1) In many potentials, specifically power-law potentials, significantly eccentric REOs do not exist above some maximum orbit frequency. In power-law potentials epicyclic theory predicts ILR orbits ($\Omega_p = \Omega - \kappa/2$) may be found at any pattern speed, but in fact only essentially circular resonant orbits are found at frequencies above the maximum. 2) At frequencies where eccentric orbits are allowed, there is a family of stable REOs with eccentricities ranging from zero up to a maximum value. This maximum eccentricity is significantly less than $1.0$, though it generally increases with decreasing orbital frequency. 3) At a given frequency, orbits with eccentricities just beyond this maximum are `sticky', i.e., stable for some cycles, and then precessing away. The greater the eccentricity, the shorter the sticking time, until at some eccentricity it is less than one orbital period, so the orbit is non-resonant. The transitions to stickiness and then instability are continuous, not abrupt. Apparently stable orbits may in fact be sticky on a very long timescale. 4) Eccentricity is a second resonance parameter beyond frequency or pattern speed for REOs. For multi-component potentials the situation may be more complicated (e.g., see Fig. 3). 

In the second part of the paper, examples were given showing how ensembles of REOs can provide the backbones for a wide variety of bars and spirals, and help explain relationships between their properties. E.g., wherever the REO families extend over a significant range of eccentricities and are excited over a small range of frequencies there are potential LBK bar backbones or backbones for slowly evolving spirals. REO ensembles of nearly constant eccentricity, and extending over range of frequencies, can produce spirals, but with more rapid wind-up. These results may generalize to three dimensional orbits, though vertical resonances will complicate the picture. This paper has focused on $x_1$ REOs moving prograde in the pattern frame. Other REO families associated with other low-order resonances exist, as will be described in later work. 

In cases where REO ensembles form wave backbones, they help us understand how: 1) bars can extend beyond the harmonic regions, 2) bars can be supported by a shell of REOs in potentials without a near harmonic region, and 3) spirals with a wide range of pattern speeds and wind-up rates can exist in different potentials. 

As illustrated in Figures 3-5 above central concentration plays a large role in determining the eccentricity range of stable REOs, and the maximum eccentricity at a given frequency. N-body simulations show that it is hard to form bars in centrally concentrated potentials. The results above suggest that this may be related to the absence of the very eccentric REOs in such cases. 

Wide, stable eccentricity ranges (and the most eccentric REOs) are generally found at low frequencies, at least in potentials like those forms studied above. This is natural because the low frequencies are associated with large circular orbits, which are located beyond central concentrations. Ensembles of such orbits can form either shell-like backbones of bars or the spirals of large radial extent. 

An external (e.g., tidal) disturbance may excite an ensemble of REOs, which form the nascent backbone of the wave and determine its pattern speed. This backbone will capture more orbits via the LBK process, and probably excite more REOs and sticky orbits. The rate of growth will depend on the initial perturbation strength and the self-gravity of the young backbone. Very strong disturbances may produce transient caustic waves, which will not adhere into a backbone in most cases, but may reappear, until phase mixed. 

N-body simulations have provided some support for the existence of a REO backbone in barred galaxies \citep{athanassoula92, athanassoula13} and spirals \citep{harsoula22}. There is also recent observational kinematic evidence for nested elliptical density contours and stellar orbits in the Large Magellanic Cloud \citep{niederhofer22}. More N-body modeling will be needed to confirm the structure of REO regions in various potentials, as well as further work to confirm the presence of nested REOs in bars and spiral waves. Testing the extended LBK growth process in N-body models is challenging. Hopefully, modelers will overcome these challenges to advance our understanding of waves in discs.

\section*{Data availability}

The numerical data generated for this article will be shared on reasonable request to the author.

\section*{Acknowledgments}

I am grateful to B. J. Smith, B. G. Elmegreen, and especially an anonymous referee for much helpful input. I acknowledge use of NASA's Astrophysics Data System.

\bibliographystyle{mn2e}

\bsp
\label{lastpage}
\end{document}